\documentclass[aps, pre, twocolumn, groupedaddress, showpacs]{revtex4-1}
\usepackage{graphicx}
\pdfoutput=1 
\usepackage{amsmath}
\usepackage{natbib}
\usepackage{color}
\usepackage{subfigure}
\usepackage{url}
\usepackage{lipsum}

\begin{document}

\title{Knudsen number dependence of 2D single-mode Rayleigh-Taylor fluid instabilities}
\author{Irina Sagert$^1$, Jim Howell$^2$, Alec Staber$^2$, Terrance Strother$^4$, Dirk Colbry$^2$, 
Wolfgang Bauer$^{2,3}$}
\affiliation{$^1$ Center for Exploration of Energy and Matter, Indiana University, Bloomington, Indiana, 47308, USA\\
$^2$Institute for Cyber-Enabled Research, Michigan State University East Lansing, Michigan 48824, USA\\
$^3$Department of Physics and Astronomy, Michigan State University, East Lansing, Michigan, 48824, USA\\
$^4$XTD-IDA, Los Alamos National Laboratory, Los Alamos, New Mexico 87545, USA}
\date{\today}
\begin{abstract}
We present a study of single-mode Rayleigh-Taylor instabilities (smRTI) with a modified Direct Simulation Monte Carlo (mDSMC) code in two dimensions. The mDSMC code is aimed to capture the dynamics of matter for a large range of Knudsen numbers within one approach. Our method combines the traditional Monte Carlo technique to efficiently propagate particles and the Point-of-Closest-Approach method for high spatial resolution. Simulations are performed using different particle mean-free-paths and we compare the results to linear theory predictions for the growth rate including diffusion and viscosity. We find good agreement between theoretical predictions and simulations and, at late times, observe the development of secondary instabilities, similar to hydrodynamic simulations and experiments. Large mean-free-paths favor particle diffusion, reduce the occurrence of secondary instabilities and approach the non-interacting gas limit. 
\end{abstract}
\maketitle
\section{Introduction}
\label{section:introduction}
Dynamical simulations that are based on interacting particles are increasingly applied in different physical fields. Different from conventional hydrodynamics methods which operate in the continuum limit, kinetic approaches are able to simulate matter at all Knudsen numbers $K = l/L$. Here, $l$ is the particle mean-free-path and $L$ is a characteristic length scale. Examples of research areas that apply kinetic methods include material science \cite{Schelling02, Kadau02}, nuclear collisions \cite{Bauer86, Bertsch84, Kruse85, Aichelin85, Aichelin86, Bouras12,Kortemeyer95} and plasma physics \cite{Casanova91, Vidal93, Vidal95}. In astrophysics, particle methods have a long history in radiation transport \cite{Roth14, Wollaeger13, Densmore04, Gentile01, Lucy99, Fleck84, Fleck71} and cosmological simulations \cite{Hernquist92}. Modern usage also includes nuclear matter in neutron star crusts \cite{Schneider13}, and neutrino transport in core-collapse supernova (CCSN) \cite{Abd12, Janka89}. Furthermore, particle methods are receiving a large interest from studies of inertial confinement fusion (ICF) capsule implosion performed at the National Ignition Facility (NIF) \cite{Casanova91,Vidal93,Vidal95,Lindl95, Lindl04, Glenzer12, Edwards13}.\\
Advantages of kinetic methods include their flexibility with regard to optical depths, the facility to include complex geometries and distributions of matter and the correct representation of the Boltzmann transport of many particles in three dimensions (3D). A current drawback in comparison to hydrodynamic simulations of macroscopic systems is the large number of particles that is required to accurately represent a physical problem and reduce statistical noise. For example, large optical depths require many particle interactions and the corresponding simulations become increasingly slow. However, as computational power increases, the relatively straight forward parallelization and scalability of particle codes \cite{Kadau06, Brunner09, Jonsson10}, could outweigh the computational costs.\\ 
Depending on the physical problem, different particle-based simulation techniques are used. Some more widely used approaches simulations include Molecular Dynamics \cite{MD_primer, Frenkel01}, Direct Simulation Monte Carlo \cite{Bird63, Bird94, Bird65} and Particle-in-Cell \cite{PIC, Hockney88}. While all approaches have been primarily developed to describe non-equilibrium matter, they are able retrieve macroscopic variables like density, fluid velocity, pressure, and temperature. Furthermore, they can model the evolution of hydrodynamic phenomena such as shock waves \cite{Bouras09, Sagert14, Li04, Gan08, Xu10, Degond10, Dimarco13} and fluid instabilities \cite{Kadau10, Kadau04, Glosli07, Weinberg14, Ashwin10, Zhakhovskii06, Zybin06, Zhakhovskii02, Nishihara10}. Both are important components in plasma and astrophysics.\\
\newline
Our goal is to develop a large-scale kinetic transport code that can handle $\gg 10^6$ particles in a computationally efficient way, and thereby simulate matter in non-equilibrium and in the hydrodynamic regime. With that, we want to study astrophysical phenomena such as core-collapse supernovae (CCSNe) \cite{Bethe90, Janka12}. Furthermore, this approach could be applied in the simulation of inertial confinement fusion (ICF) capsule dynamics \cite{Lindl95, Lindl04, Glenzer12, Edwards13}. The evolution of both systems, CCSNe and ICF, is largely governed by shock wave dynamics paired with fluid instabilities and non-equilibrium particle transport \cite{Bodner74, Rosenberg14, Hurricane14, Colgate66, Janka12, Ott13, Burrows13, Sumiyoshi12, Bruenn13}.\\
Our modified Direct Simulation Monte Carlo code (mDSMC) has already been successfully tested on shock wave phenomena in non-equilibrium and in the continuum regime \cite{Sagert14, Sagert13}. In this work we present our first detailed study of fluid instabilities in two dimensions (2D). Here, we focus on the single-mode Rayleigh-Taylor instability (smRTI) \cite{Liska03}. Our motivation is the possible important role of RTIs in ICF and CCSN dynamics. The advantage of the smRTI is that at early stages, it can be compared to an analytic solution from linear theory. With that, we can refer to the latter and experiments for comparison. Fluid instability simulations including RTIs have been performed by particle codes in the past (see e.g. \cite{Kadau10, Gan11}). In this paper, we present a detailed and comprehensive analysis for a large range of particle mean-free-paths.\\ 
\newline
In the following, we give a short introduction of RTIs in section \ref{section:RTI}, followed by an overview of our mDSMC code in section \ref{section:approach}. We then proceed with our simulation setup and discuss the results for varying particle mean-free-paths in sections \ref{section:simulation_setup} and \ref{section:simulation_results}. The paper closes with a summary and outlook in section \ref{section:conclusion}.
\section{Rayleigh-Taylor instabilities} 
\label{section:RTI}
Rayleigh-Taylor instabilities form at the interface of two fluids with different densities when the less dense fluid is pushing against the one with higher density. A typical example is a heavy fluid resting on top of a lighter one in the presence of a gravitational acceleration \cite{Rayleigh82, Taylor50}. In such a case, small perturbations at the fluid interface grow and result in the development of RTIs. The latter can be found in many different physical environments - ranging from astrophysical systems to geophysical phenomena. Due to their large dynamical impact and the direct connection with turbulent mixing, RTIs receive wide interest. Studies have been performed analytically, experimentally and numerically \cite{Sharp84, Kull91, Wei12}, while the growing computational power allows to study RTI phenomena in greater detail and for increasingly larger systems.\\
In realistic environments, fluid instabilities of many different wavenumbers can be present. For code validation studies, it is easier to focus on the smRTI which arises from an initial perturbation $\eta_0 (x)$ with a defined wavelength $\lambda$. Its evolution can be divided into several major stages: \\
During the first stage, the amplitude of the initial perturbation is $2\:B \lesssim 0.5 \lambda$. Here, the instability undergoes exponential growth which can be described by linear analysis \cite{chandra,Frieman54}:
\begin{align}
\eta( x, t) = 0.5 \left( e^{\gamma t} + e^{-\gamma t} \right) \eta_0 (x). 
\label{eta} 
\end{align}
For ideal fluids, the growth rate $\gamma$ is given by:
\begin{align}
\gamma = \sqrt{Ag k}, \: \: \: A = \frac{\rho_2 - \rho_1}{ \rho_2 + \rho_1}, \: \: \: k = 2 \pi /  \lambda \: , 
\label{gamma_ideal}
\end{align}
$A$ being the Atwood number and $k$ the wavenumber. The densities of the light and heavy fluids are given by $\rho_1$ and $\rho_2$, respectively. When diffusion and viscosity are included, the growth rate changes to:
\begin{align}
\gamma(t) = \sqrt{ \frac{A g k}{\phi(A, t)} + \nu^2 k^4} - (\nu + \xi)k^2 \: .
\label{gamma_visc}
\end{align}
Here, $\phi(A, t)$ represents dynamic diffusion effects, $\xi$ is the diffusion coefficient, and $\nu$ the kinematic viscosity. Note that, different from the ideal fluid approximation in eq.(\ref{gamma_ideal}), the viscous growth rate is dependent on time $t$.\\ 
When $B > \lambda$, non-linear effects start to dominate. Light fluid bubbles rise into the denser phase, while spikes of the latter sink downwards. Perturbations with large wave numbers are generated and aerodynamical deceleration of sinking spikes leads to the formation of mushroom shaped jets \cite{Inogamov99}. As bubbles and spikes start to interact with each other, the dynamics become chaotic, leading to turbulent mixing.\\
The effects that compressibility, viscosity and surface tension have on the evolution of the RTI have been discussed in e.g. \cite{Livescu04, Ribeyre04}. Here, surface tension and viscosity were found to stabilize perturbations while simulations using compressible fluids experience delays in the formation of the mushroom shaped jets. Challenges in numerical studies of RTIs arise in the non-linear regime when a finer computational grid leads to the development of more secondary instabilities. This is partly due to the finer resolution and partly due to the use of a different grid. Convergence tests are generally performed to ensure that the dynamics of the simulated system do not change with resolution.   
\section{Modified DSMC approach}
\label{section:approach}
Here we present a short overview of our simulation method adjusted to 2D calculations. A general discussion can be found in \cite{Sagert14}. The foundation of our approach is a DSMC method where the phase-space of the physical problem is represented by $N$ $\delta$-functions or so-called \textit{test-particles}:
\begin{equation}
f(\mathbf{r},\mathbf{p},t)=\sum_{i=0}^{N} \delta^{2}\big(\mathbf{r}-\mathbf{r}_i(t)\big)\delta^{2}\big(\mathbf{p}-\mathbf{p}_i(t)\big) .
\end{equation}
Here, $\mathbf{r}_i$ is the position and $\mathbf{p}_i$ the momentum of the $i$th test-particle. This distribution function is used as input into the Boltzmann equation \cite{Chapman70} and results in $2 N$ ordinary differential equations of motion for each test-particle with mass $m_i$:
\begin{align} 
\frac{d}{dt} \mathbf{p}_i &= \mathbf{F}(\mathbf{r}_i)+ \mathbf {\cal C}(\mathbf{p}_i), \\
\frac{d}{dt} \mathbf{r}_i &= \frac{\mathbf{p}_i}{m_{i}}, \: \: \: i = 1,\ldots,N .
\label{eom}
\end{align}
Particles interact with each other via one-body mean-field forces $\mathbf{F}(\mathbf{r}_i)$, such as gravity. In addition, they undergo two-body interactions which are symbolized by $\mathbf{\mathcal{C}}(\mathbf{p}_i)$. For realistic fluids, the latter must contain the appropriate cross-section $\sigma$. In the current study, we want to test the continuum behavior of our code. Particle interactions are therefore modeled as simple elastic collisions. More complex interactions will be implemented in the future, as has already been done in previous works for CCSNe simulations \cite{Strother10, Strother07, Bauer05}. For elastic collisions, the 2D cross-sections are related to a particle effective interaction radius $\sigma_\mathrm{2D} = 2 \: r_\mathrm{eff}$. Here, $r_\mathrm{eff}$ is determined by the particle mean-free-path $l$ and the number density $n$ which is defined as the number of particles $N$ divided by the area $A$:
\begin{equation}
r_{\mathrm{eff}} = 1/(2 \: l \: n), \:\:\: n=N/A \: . 
\label{reff_l}
\end{equation}
As in our previous works, $l$ will be used as an input variable. From that, we determine the particle effective radii and apply them in our collision partner search \cite{Sagert14}. Hereby, we calculate the time $t_o$ at which the effective radii of both particles overlap: 
\begin{align}
t_{o (1,2)} &= t_p \pm \sqrt{ t_p^2 -  t_\mathrm{rel}^2 + t_\mathrm{eff}^2 } \: , 
\label{collision_time2} \\
t_p &= - \left( \mathbf{r}_{ \mathrm{rel} } \cdot \mathbf{v}_{ \mathrm{rel} } \right) /  \left| \mathbf{v}_{ \mathrm{rel} } \right|^2 ,
\label{tp}\\
t_\mathrm{rel} &= |\mathbf{r}_{\mathrm{rel}} | /   \left| \mathbf{v}_{ \mathrm{rel} } \right| ,\\
t_\mathrm{eff} &= ( r_{\mathrm{eff}, A} + r_{\mathrm{eff}, B} ) /  \left| \mathbf{v}_{ \mathrm{rel} } \right| \: .
\end{align}
If either $t_{o1}$ or $t_{o2}$ is a real number, a collision can take place \cite{Sagert14}. Otherwise, the particles are too far away from each other. Note, that the effective radii are utilized in this step only. For the actual interaction time we choose the Point-of-Closest-Apprach (PoCA) method which is different from the usual DSMC routine. Here, the collision is performed at $t_p$, when two particles reach their minimal distance. With that, the PoCA algorithm reduces causality violations, which are often present in DSMC type simulations \cite{Kortemeyer95}. The combination of DSMC and PoCA results in a favorable scaling of the computational time with $N$ \cite{Sagert14, Howell13} and an improved spatial accuracy. When the collision is performed at the point of closest approach, the outgoing particle velocity vectors are determined randomly in the center-of-mass (cm) frame of the colliding pair:
\begin{align}
 \phi &= 2 \pi \kappa, \: \kappa \in [0.0, 1.0]\: , \\
v_{x, \mathrm{out, cm}} &=  v_\mathrm{in,cm} \cos( \phi) \:, \\
v_{y, \mathrm{out,cm}} &=  v_\mathrm{in,cm} \sin( \phi) \:, \\
v_\mathrm{in,cm} &= \sqrt{ v^2_{x, \mathrm{in,cm}} + v^2_{y, \mathrm{in,cm}} } \: . 
\end{align}
From the cm frame they are then transformed back into the laboratory frame. \\
At the beginning of each time step, particles are sorted into their spatial cells or \textit{bins} where they can only interact with partners in their own bin or adjacent cells. To prevent particles from traveling beyond their collision neighborhood during a time step $\Delta t$, the latter is given by the cell size $\Delta x$ divided by the maximum particle velocity:
\begin{equation}
\Delta t (t) =  \Delta x / v_{ \mathrm{max}  (t) }. 
\end{equation}
We use Euler's method to iterate the equations of motion: 
\begin{align}
\mathbf{r_i}(t + \Delta t) &= \mathbf{r_i}(t) + \mathbf{v}_{\mathrm{old},i} \: t_{\mathrm{p}} + \mathbf{v}_{\mathrm{new},i} 
( \Delta t - t_{\mathrm{p}})
\label{ru_c}\\
\mathbf{r_i}(t + \Delta t) &= \mathbf{r_i}(t) + \mathbf{v}_{\mathrm{old},i} \: \Delta t \: ,
\label{ru_nc} 
\end{align}
whereas eq.(\ref{ru_c}) is applied when a collision takes place and eq.(\ref{ru_nc}) otherwise. Computationally expensive sorting algorithms are not necessary as we represent each grid cell via a linked list of particles it contains which only requires a simple coordinate check.\\
Our code can perform simulations in 2D and 3D whereas the degrees of freedom and matter equation of state change accordingly. The collision partner search is parallelized using shared memory parallelization via OpenMP. However, our 2D RTI studies are long-time simulations. Using OpenMP with 32 processors they take on average $\sim 1.2 \times 10^5$ time steps in $\sim 350\:$ hours. To enable larger 2D and 3D simulations in the future, a distributed memory parallelization via MPI is necessary and is currently under development \cite{Howell13}. The scaling of the collision partner search has been tested in the current OpenMP and preliminary MPI setups and is close to ideal for 2D and 3D simulations \cite{Sagert14, Howell13}. In general, our study of the smRTI uses the same algorithm as in \cite{Sagert14}, however, we include a change in the boundary conditions for the long-time evolution of the RTI (see next section).  
\section{Simulation Setup}
\label{section:simulation_setup}
\subsection{Particle initialization}
The 2D smRTI is initialized as a heavy fluid with density $\rho_2$ lying on top of a light one with $\rho_1 = 0.5 \rho_2$. The box size is $0 \leq x \leq L_x$ and $0 \leq y \leq L_y$. Both fluids are initially at rest with a pressure $P_0$ at the fluid interface. The units of all quantities are given by the dimensions of length $\tilde{L}$, density $\tilde{\rho}$, and pressure $\tilde{P}$. Consequently, the units for velocity are $\sqrt{\tilde{P} /  \tilde{\rho}}$, the units for time $\tilde{L} \: \sqrt{\tilde{\rho} /  \tilde{P}}$, and the units for the gravitational acceleration $ \tilde{P} / ( \tilde{L}  \tilde{\rho})$. In the latter, g is set to 1.0 and is pointing in the negative $y$-direction. The simulation space is divided into two equally sized areas $A_1 = A_2$ whereas $A_2$ contains the high-density fluid while the low-density fluid is in $A_1$. Both sub-spaces contain the same number of test-particles $N_1 = N_2$ with masses $m_{1,2} =  \rho_{1,2} A_{1,2} / N_{1,2}$. To keep track of their motion, particles in $A_2$ are assigned a \textit{particle type} $\tau_2 = 2$ while particles $A_1$ are given $\tau_1 = 1$. \\
The pressure as a function of the $y$-position is given by the barometric formula. Assuming that the densities $\rho_1$ and $\rho_2$ are constant and do not depend on the height, the expression for the pressure is given by:
\begin{equation}
P_{1,2} (y) = P_0 + \rho_{1,2} \: g \: (y - 0.5\:L_y)\:, 
\label{pressure}
\end{equation}
whereas we choose $P_0 = 2.5$. This allows the determination of the temperature at height $y$ via the ideal gas law:
\begin{equation}
k\:T_{1,2}(y) = P_{1,2}(y)/n_{1,2} \: .
\end{equation}
With given $P$, $k\:T$, and their connection to the the root-mean-square velocity $P = v^2_{\mathrm{rms}} \:n m /2$, we can initialize absolute velocities for each particle $i$ at $y_i$ with random directions by a 2D Maxwell-Boltzmann (MB) distribution:
\begin{equation}
F_{2D,v} (v, y_i) = \left( \frac{m}{2\pi kT(y_i)}\right) \exp\left( - \frac{mv^2}{kT(y_i)} \right)
\end{equation}
and a Monte-Carlo algorithm \cite{Sagert14b}. Furthermore, to initialize the smRTI, a perturbation $\eta_0(x)$ is introduced via a modification to the fluid interface: 
\begin{equation}
\eta_0(x) = 0.5 L_y + B_0 \cos(2 \pi x / \lambda)
\label{eta_0}
\end{equation}
with an amplitude $B_0 = 0.01$ and wavelength $\lambda = 2 \: L_x$. 
\subsection{RTI growth rate with $D$ and $\nu$}
\label{section:gamma_D_nu}
In 2D systems, expressions for the dynamic viscosity $\mu$ and diffusion coefficient $\xi$ can be determined from kinetic theory (using the approach of \cite{Plawsky}):
\begin{equation}
\mu = \rho\: \bar{v} l /2, \: \: \: \xi =  \bar{v} l/2 .
\end{equation}
Here, $\bar{v}$ is the 2D mean velocity 
\begin{equation}
\bar{v} = \sqrt{kT/(2m)}
\label{velocities}
\end{equation}
and the kinematic viscosity $\nu$ in eq.(\ref{gamma_visc}) can be obtained by $\nu = \mu/\rho = \xi$. To determine $\phi(A,t)$ in eq.(\ref{gamma_visc}), we apply the approach of \cite{Duff62} and numerically solve the following eigenvalue equation for the vertical velocity component $\omega$:
\begin{equation}
a^2 (t) \frac{d}{d \sigma} \left[ \Psi \frac{d\omega}{d\sigma} \right] = \omega \left( \Psi - a(t) \phi \: \frac{dQ}{d\sigma} \right),
\label{eigenvalue}
\end{equation} 
with:
\begin{align}
\Psi & =  1 + AQ,\:\: a (t) = 1/(k \epsilon),\: \: \epsilon = 2 \sqrt{\xi t}, \\
Q(\sigma) &= 2 \pi^{-1/2} \int_0^\sigma \exp( - \zeta^2) d \zeta = \mathrm{erf}(\zeta)
\end{align}
and $\sigma$ being the scaled vertical direction $(y - 0.5\: L_y)/\epsilon$. The boundary conditions are $\omega \rightarrow 0$ for $\sigma \rightarrow \pm \infty$. For the solution, eq.(\ref{eigenvalue}) is replaced by a finite-difference analogy. The value of $\omega (\sigma + \delta \sigma)$ can be obtained from the knowledge of $\omega(\sigma)$ and $d \omega (\sigma) / d \sigma$ via iterations. These are started at $\sigma \ll -1$, so that $\omega (\sigma) = \exp(\sigma/a)$ can be assumed. For trial values of $\phi$ and assigned $A$ and $t$, the solution for $\omega$ is obtained to sufficiently large $\sigma$. For the correct $\phi$, $\omega$ should approach zero, and otherwise diverge to $+ \infty$ or $- \infty$. We apply a root-finding routine to determine the value of $\phi$. Due to the time-dependence of the growth rate $\gamma(t)$, the evolution of the instability is now given by:
\begin{align}
\eta (x,t) &= \Gamma(t) \: \eta_0 (x), \\
\Gamma(t) &= 0.5 \left( e^{\beta(t)} + e^{-\beta(t)} \right), \: \beta(t) =  \int_0^t \gamma(t) dt,
\label{Gamma}
\end{align}
whereas $\beta(t)$ is determined by numerical integration.
\begin{figure*}
\centering
\subfigure{
\includegraphics[width=0.42\textwidth]{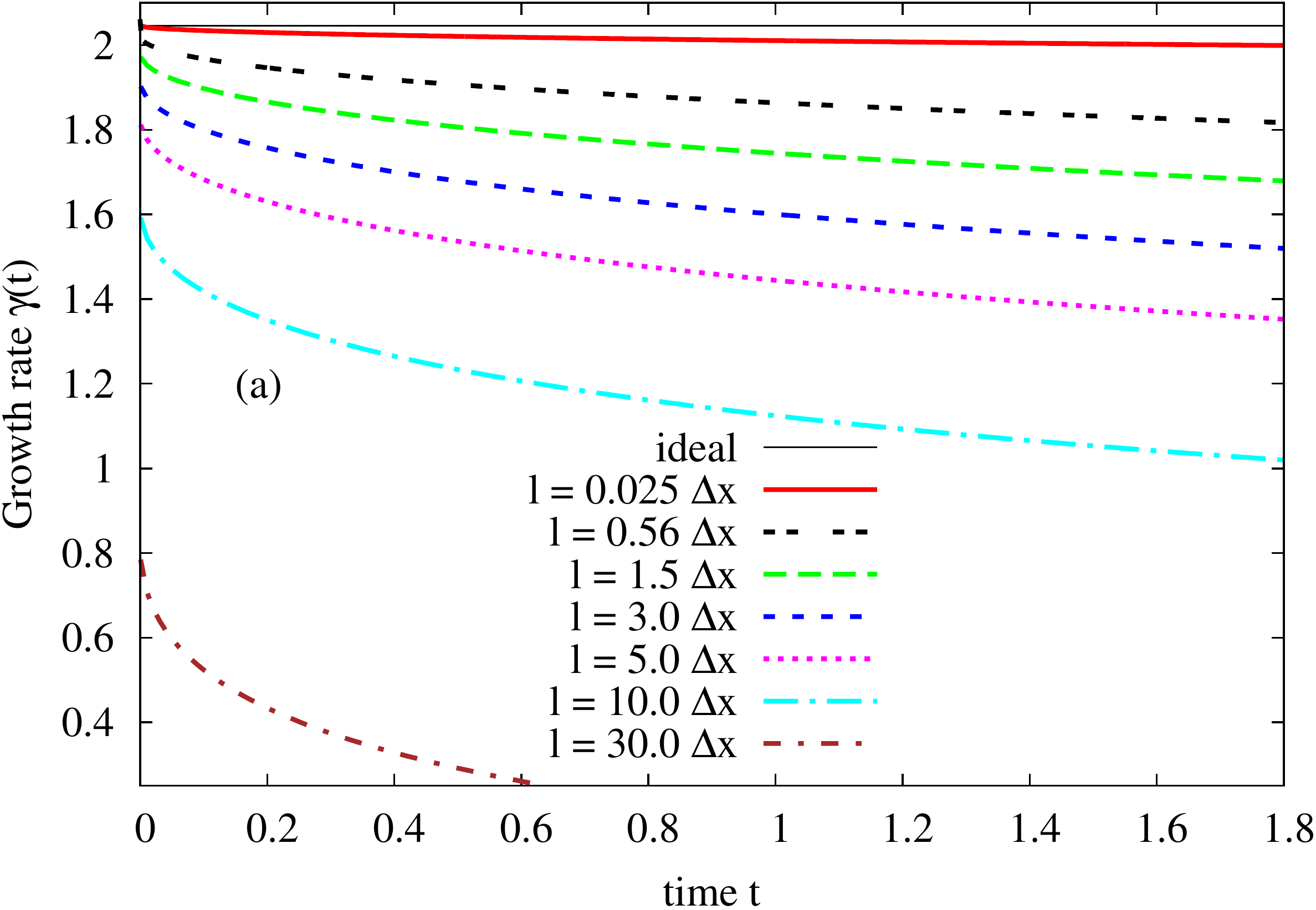}}
\subfigure{
\includegraphics[width=0.42\textwidth]{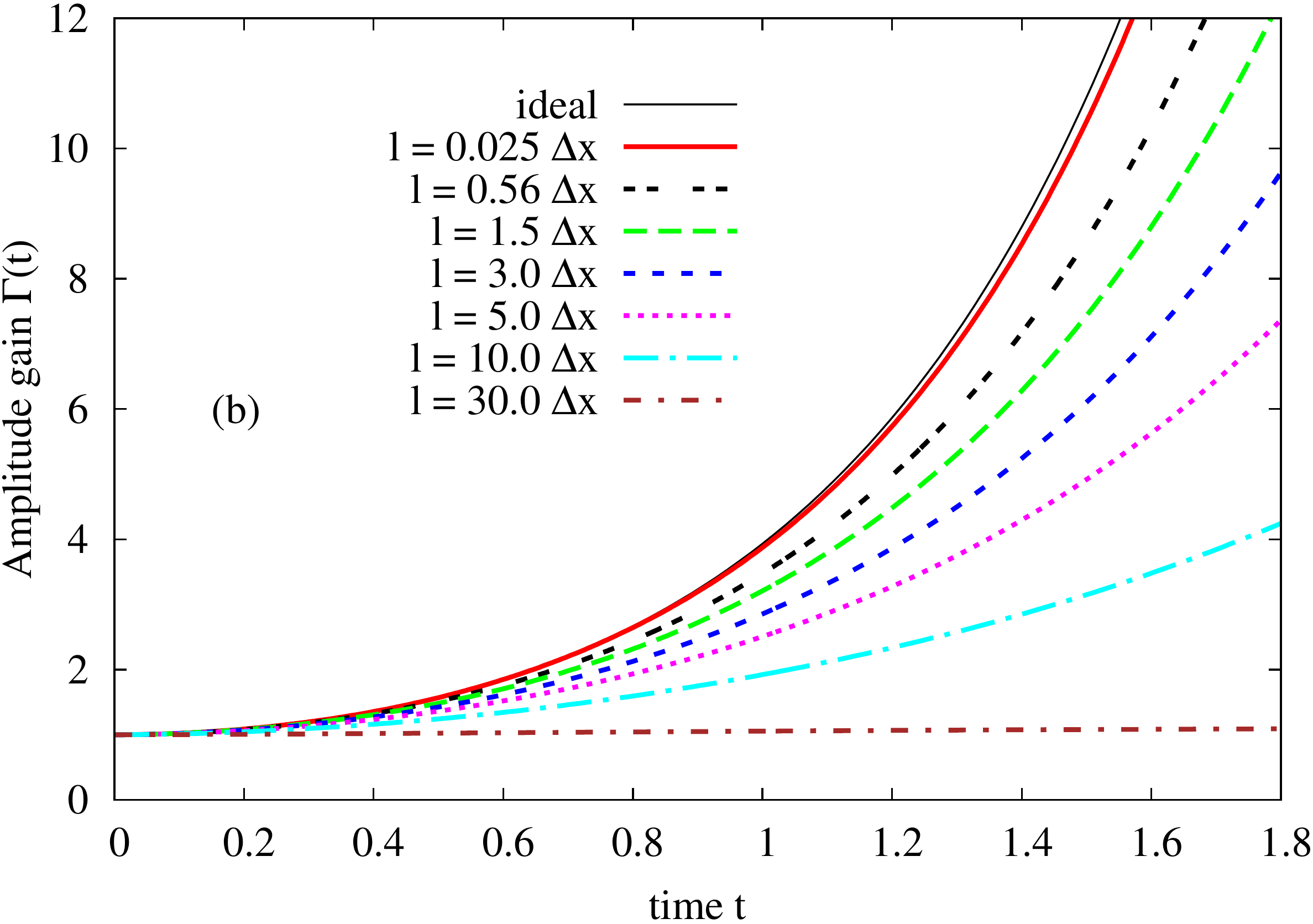}}
\caption{(Color online) (a) Growth rate $\gamma(t)$ and (b) amplitude gain $\Gamma(t)$ for different mean-free-paths $l$ according to eq.(\ref{gamma_visc}) and eq.(\ref{Gamma}). The values of $\gamma$ and $\Gamma$ in the limit of ideal hydrodynamics are given by thin solid lines.}
\label{Fig_gamma}
\end{figure*}
Fig.\:\ref{Fig_gamma} shows the values of $\gamma (t)$ and the amplitude gain $\Gamma(t)$ for $l = 0.025\:\Delta x - 100.0\:\Delta x$. The growth rate is almost constant for $l = 0.025 \: \Delta x$ and close its ideal fluid value of $\gamma \sim 2.046$. For larger mean-free-paths, viscosity and diffusion lead to a decrease in $\gamma(t)$. For $l = 100.0\:\Delta x$, we find that $\gamma < 0$ at all times, which implies that the smRTI evolution is dominated by diffusive effects from the very beginning \cite{Duff62, Wei12}. In this case, an instability will not develop and the two gases will diffuse into each other. Table\:\ref{table1} lists the exact values of $\Gamma(t)$ for $t=0.5, \: 1.25$, and $1.75$ together with the kinematic viscosities $\nu$, Reynolds numbers \cite{Wei12}
\begin{equation}
R = \lambda \nu^{-1} \sqrt{A \left(A+1\right)^{-1} g \lambda} 
\end{equation}
\begin{table*}
\begin{tabular} {  c  | c | c | c | c | c | c | c   }
\hline
$l \: [\Delta x]$ & $\nu \: [\sqrt{\tilde{P}/ \tilde{\rho}} \: \tilde{L} \: ] $ & $\Gamma(t = 0.5)$ & $\Gamma(t = 1.25)$ & $\Gamma(t = 1.75)$ & $\gamma_\mathrm{nd}$ & $\Gamma_\mathrm{nd} (t=1.75)$ & $R$ \\
\hline
\hline
 0.025  &  $3.515 \times 10^{-6}$  & 1.559 & 6.338 & 17.156 &  2.025 & 17.313 & $5.030 \times 10^5$ \\
 0.56    &  $7.873 \times 10^{-5}$  & 1.515 & 5.457 & 13.548 & 1.934 &  14.769 & $2.245 \times 10^3$ \\
 1.5       &  $2.109 \times 10^{-4}$  & 1.485 & 4.881 & 11.322 & 1.852 &  12.800 & $8.383 \times 10^2$ \\
 3.0       &  $4.217 \times 10^{-4}$  & 1.426 & 4.172 &   8.931 & 1.759 &  10.883 & $4.192 \times 10^2$ \\
 5.0       &  $7.029 \times 10^{-4}$  & 1.363 & 3.508 &   6.891 & 1.661 &    9.176 & $2.515 \times 10^2$ \\
 10.0     &  $1.406 \times 10^{-3}$  & 1.244 & 2.456 &   4.040 & 1.471 &    6.599 & $1.257 \times 10^2$ \\
 30.0     &  $4.217 \times 10^{-3}$  & 1.024 & 1.069 &   1.089 & 1.005 &    2.989 & $4.192 \times 10^1$ \\
\hline
\end{tabular}
\caption{Kinematic viscosities $\nu$, amplitude gain with and without diffusion $\Gamma (t)$ (see text) and $\Gamma_\mathrm{nd} (t)$  \cite{chandra}, respectively, growth rates without diffusion $\gamma_\mathrm{nd}$, and Reynolds numbers $R$ for different particle mean-free-paths $l$.}
\label{table1}
\end{table*}
and non-diffusive viscous growth rates $\gamma_\mathrm{nd}$ and $\Gamma_\mathrm{nd}$ \cite{chandra}. The latter are similar to $\Gamma$ but generally higher which leads to a faster smRTI evolution. For comparison with our simulations we will use the diffusive viscous growth rates $\Gamma(t)$.
\subsection{Boundary conditions}
Our previous studies apply simple reflective boundary conditions \cite{Sagert14}. For example, if during $\Delta t$ a particle with velocity $\mathbf{v}_{\mathrm{old}}$ crosses the box boundary at $y = b_y = 0$ to a position beyond the boundary $\mathbf{r}_{\mathrm{old}}$, its location and motion are updated via: 
\begin{align}
x_{\mathrm{new}} &=  x_{\mathrm{old}}, \: \:  y_{\mathrm{new}} = - y_{\mathrm{old}}\:,\\
v_{\mathrm{new},x} &= v_{\mathrm{old},x}, \: \: v_{\mathrm{new},y} = - v_{\mathrm{old},y}\:.
\end{align} 
For the current tests, we modify this simple approach: \\
In addition to the usual particle motion, we have to consider that the gravitational acceleration $g$ is present at all times. When a particle is moving towards $b_y=0$, it is accelerated downwards. Once it is reflected by the wall and moves in the opposite direction, its $y$-velocity is decreased due to $g$. Furthermore, with typically $\mathcal{O}(10^5)$ simulation time steps, the sinusoidal form of the smRTI instability together with the simple reflective boundary conditions could lead to the development of standing waves or shock waves in the simulation box. Since these could impact the evolution of the smRTI, we modify the boundary conditions so that particles which interact with the walls receive a random new direction of motion. We refer to these as \textit{random reflective} boundary conditions.\\
To implement these modification, we determine the particle-boundary collision time $t_b$ from: 
\begin{align}
b_y &= y_\mathrm{old} + v_{\mathrm{old},y} \: t_b +  g \: t^2_b / 2,\\
t_b  &= \frac{1}{g} \left( -v_{\mathrm{old},y}  \pm \sqrt{ v^2_{\mathrm{old},y}  - 2 \:( y_{\mathrm{old},y} - b_y)} \: \right),
\end{align}
whereas $b_y = L_y$ or $b_y = 0$. The incoming $y$-velocity at $b_y$ is given by:
\begin{equation}
v_{b,y} = v_{\mathrm{old},y} + g \: t_b \:\:. 
\end{equation} 
Together with the unchanged particle motion in the $x$-direction, the absolute velocity $v_b$ at $b_y$ becomes: 
\begin{equation} 
v_b = \sqrt{v^2_{b,y} + v^2_{x, \mathrm{old}}} \:\:.
\end{equation}
To determine the post-reflection velocity, the outgoing direction of motion is randomized. For that, we create random numbers $\kappa_x \in[-1.0, 1.0]$ and $\kappa_y \in [0.0, 1.0]$ or $\kappa_y \in [-1.0, 0.0]$, depending on whether the reflection is performed at $b_y = 0$ or $b_y = L_y$, respectively. The random numbers are scaled so that $\kappa_x^2 + \kappa_y^2 = 1$, and the new velocity components and positions calculated as:
\begin{align}
v_{\mathrm{new},y} &=  v_b \: \kappa_y + g \: (\Delta t - t_b),\\ 
v_{\mathrm{new},x} &= v_b \: \kappa_x, \\
y_{\mathrm{new},y} &= b_y +  v_b \: \kappa_y \: (\Delta t-t_b) + g \: (\Delta t-t_b)^2 / 2, \\
x_{\mathrm{new},x} &= x_\mathrm{old} + v_{\mathrm{old},x} \: t_b + v_{\mathrm{new},x} \: (\Delta t - t_b).
\end{align}
With that, our simulation should be able to disperse incoming waves and shocks at the boundaries and thereby minimize wall effects.
\subsection{Minimal mean-free-path}
\label{section:minimal_l}
In our previous studies \cite{Sagert14}, we set the mean-free-path to small values of e.g. $l = 10^{-3}\:\Delta x$ to simulate matter in the continuum regime. This was motivated by the effective particle radius being not only dependent on $l$ but also on the particle number per bin $N = N_\mathrm{bin}$ as is shown in eq.(\ref{reff_l}) with $A = A_\mathrm{bin} = \left( \Delta x \right)^2$. In our shock wave studies, $N_\mathrm{bin}$ varies up to a factor $\sim 50$. In this case, setting $l$ to a small value $l \ll \Delta x$ ensures that the effective radii stay sufficiently large for particles to see all potential interaction partners in the collision neighborhood. For the smRTI, the number of particles per calculation bin fluctuates around $N_\mathrm{bin} \sim 10$. Considering, that a particle can only detect collision partners in its collision neighborhood we can set a maximal limit on the effective radius to $r_\mathrm{eff, max} \sim 2 \Delta x$ (considering two interacting particles at opposite corners of the collision neighborhood) which results in a minimal value of the mean-free-path:
\begin{equation}
l_{\mathrm{min,1}} = \Delta x/(4 \: N_{\mathrm{bin}}) \sim 0.025 \: \Delta x.
\label{lmin}
\end{equation}
Although for $l \ll l_\mathrm{min,1}$, the effective radii technically increase, particles are still unable to see beyond the collision neighborhood. While all simulations with $l \leq l_\mathrm{min,1}$ should therefore evolve similarly, comparisons with theoretical predictions that involve viscosity and diffusion should consider the true minimal value of $l$. As seen from eq.(\ref{lmin}), the only possibility to reduce $l_\mathrm{min,1}$ is to increase the value of $N_\mathrm{bin}$ keeping $\Delta x$ constant, or decrease the latter at a constant $N_\mathrm{bin}$. The value of $l_\mathrm{min}$ might even increase if we assume that in the continuum limit all particles interact and that collisions generally take place between close neighbors. In this case, we can determine $r_\mathrm{eff}$ from a particle area: 
\begin{align}
A_p &= \pi \: r^2_p   = A_\mathrm{bin} / N_\mathrm{bin},\\
\rightarrow r_p &= \sqrt{A_\mathrm{bin} / (\pi \: N_\mathrm{bin}) } = r_\mathrm{eff} .
\label{rp}
\end{align}
Again, in principle, a larger effective radius enables particles to see beyond their closest collision partner, however, the interaction will still take place between particles with a distance $\leq 2\: r_\mathrm{eff}$. Using eq.(\ref{reff_l}) and eq.(\ref{rp}) we arrive at a new mean-free-path limit of $l_\mathrm{min,2} = \sqrt{ \pi / (4 N_\mathrm{bin})} \: \Delta x \sim 0.28 \: \Delta x$. Furthermore, in our collision algorithm, we discard interactions that involve more than two particles. This situation occurs frequently for close-to-continuum simulations and, as a result, about half of potential collisions are not performed when $l$ is small \cite{Sagert14}. To account for the fact that we have about 50\% fewer collisions during a time step than anticipated and therefore more free streaming particles, we increase the minimal mean-free-path to 
\begin{equation}
l_\mathrm{min,2} = 2 \times 0.28 \: \Delta x = 0.56 \: \Delta x. 
\end{equation}
In the following we will use $l=l_\mathrm{min,1}$ for our close-to-continuum simulations but also apply $l_\mathrm{min,2}$ when comparing results with theoretical predictions.
\section{Mean-free-path studies}
\label{section:simulation_results}
\subsection{Evolution of a plane fluid interface}
Before discussing the smRTI simulations, we will study the evolution of a plane fluid interface with $B_0 = 0$. Different from hydrodynamic codes, the finite number of particles in kinetic methods always leads to small irregularities of the fluid interface. If not suppressed, these serve as seeds for the development of large-scale instabilities. 
\begin{figure*}
\begin{center}
\includegraphics[width = 0.90\textwidth]{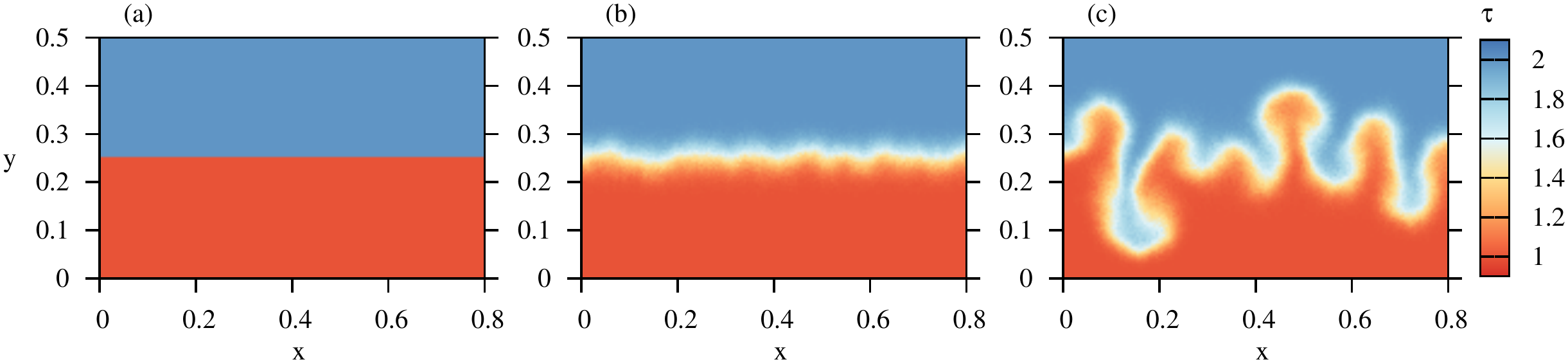}
\caption{(Color online) Evolution of fluid instabilities at an initially unperturbed interface of a high density (blue, above white separation interface) and a low density (red, below white separation interface) fluid in the presence of a gravitational acceleration. The particle number is $N =10^7$. We plot the particle type $\tau$ at times (a) $t=0$, (b) $t=1.0$, and (c) $t=3.0$.}
\label{RT_pert}
\end{center}
\end{figure*}
Fig.\ref{RT_pert} shows the evolution of the initially smooth interface for $t \leq 3.0$ using $10^7$ test-particles. The initialization of the simulation is as in section \ref{section:simulation_setup}, however the box size is chosen as $L_x = 0.8$ and $L_y = 0.5$ with $1280 \times 800$ calculation bins and $160 \times 100$ output bins. The boundary conditions are random reflective and particles are interacting with each other via simple elastic collisions according to the PoCA algorithm using $l = l_\mathrm{min,1}$.\\ 
It can be seen from Fig.\:\ref{RT_pert} that the seemingly smooth fluid interface develops a diffusion layer with small dips and peaks within $t \ll 1.0$. This mixing is due to the finite value of $l$ that allows particles to move from one fluid into the other. The peaks and dips are caused by irregularities of the fluid interface as a consequence of the random initial particle positioning. As the perturbations grow over time they result in the formation of large scale RTIs with different wavelengths. We find a similar behavior in our close-to-continuum smRTI simulation that we discuss in the next section. It is important to point out that a small modification of the particle initialization via e.g. changed particle positions, can results in a different RTI evolution at later times.
\subsection{smRTI close to the continuum limit}
\label{subsection:smRTI_continuum}
\begin{figure*}
\begin{center}
\includegraphics[width = 0.85\textwidth]{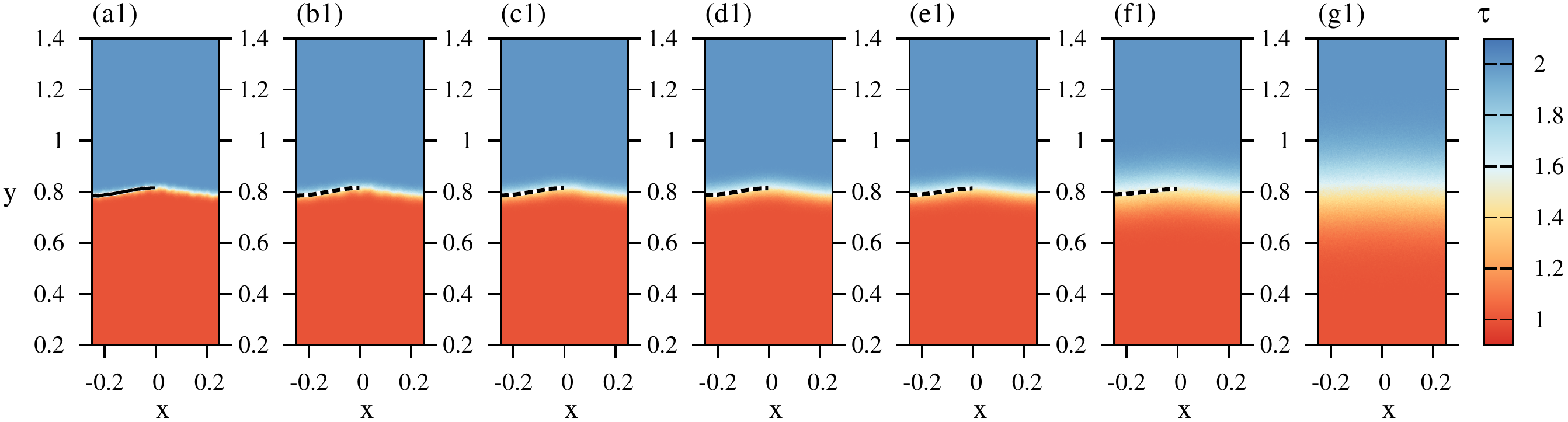}
\includegraphics[width = 0.85\textwidth]{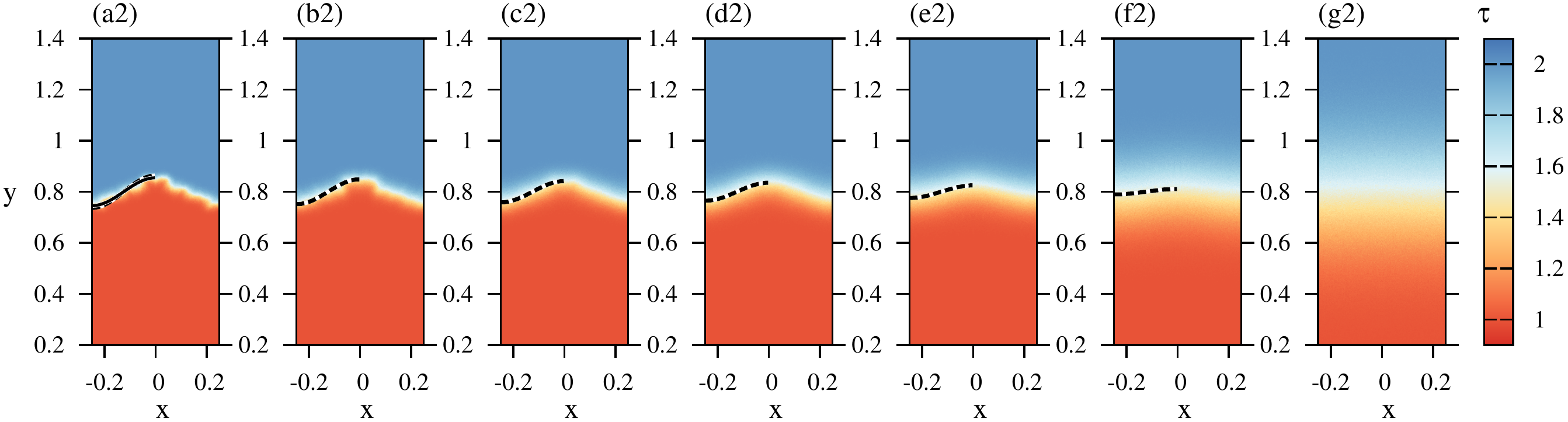}
\includegraphics[width = 0.85\textwidth]{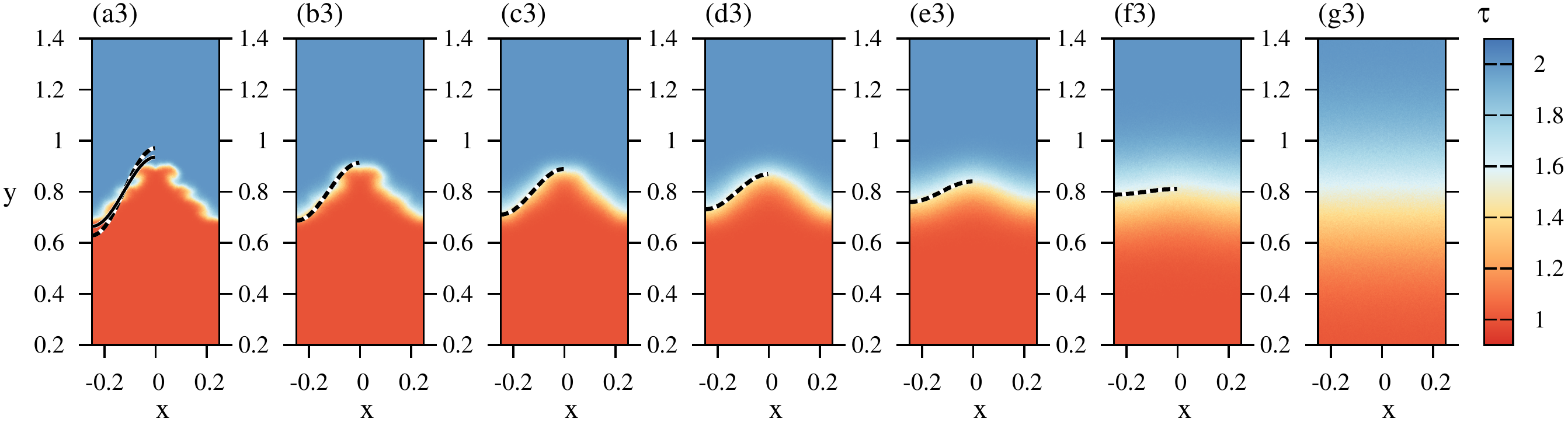}
\includegraphics[width = 0.85\textwidth]{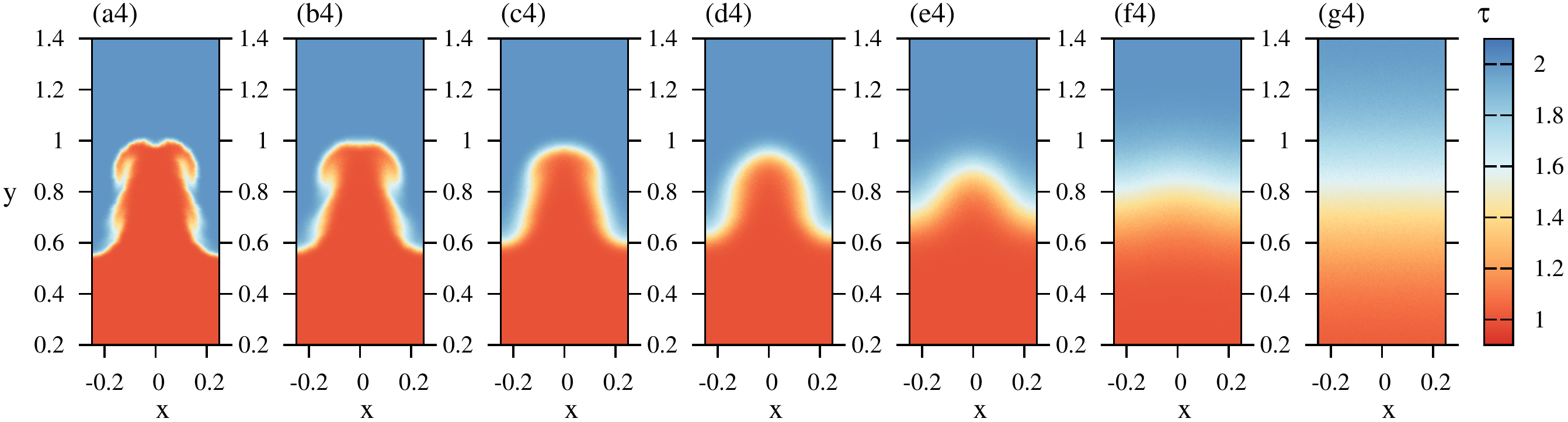}
\includegraphics[width = 0.85\textwidth]{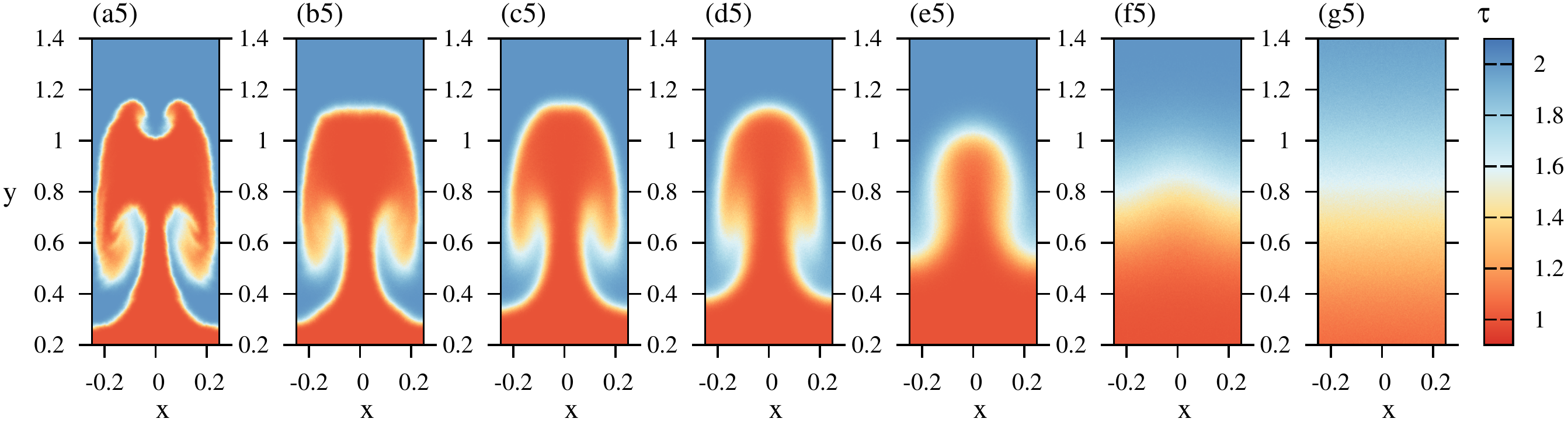}
\caption{(Color online) Time evolution of the average particle type $\tau$ per bin in the smRTI simulation for: (a) $l=0.025 \: \Delta x$, (b) $l=1.5 \: \Delta x$, (c) $l=3.0 \: \Delta x$, (d) $l=5.0 \: \Delta x$, (e) $l=10.0 \: \Delta x$, (f) $l=30.0 \: \Delta x$, and (g) $l=100.0 \: \Delta x$. Snapshots are taken at times (1) $t=0.5$, (2) $t=1.25$, (3) $t=1.75$, (4) $t=2.5$, and (5) $t=3.75$. Dashed lines give the analytic predictions using the corresponding growth rates for $l$ from table \ref{table1}. The solid lines in subfigures (a) are the analytic solution for $l = l_\mathrm{min,2} = 0.56 \Delta x$. The number of test-particles is $N = 4.0 \times 10^7$ with $800 \times 5120$ simulation bins covering a space of $0.25 \times 1.6$.}
\label{RT_mfp}
\end{center}
\end{figure*}

We start with a smRTI simulation close to the hydrodynamic limit by setting $l = l_\mathrm{min,1} = 0.025 \: \Delta x$. The box dimensions are $L_x = 0.25$ and $L_y = 1.6$ with $800 \times 5120$ calculation bins, and $100 \times 640$ output bins. The width of a calculation bin is thereby $\Delta x = 3.125 \times 10^{-4}$. We use $N = 4.0 \times 10^7$ test-particles.\\
\begin{figure}
\begin{center}
\includegraphics[width = 0.4\textwidth]{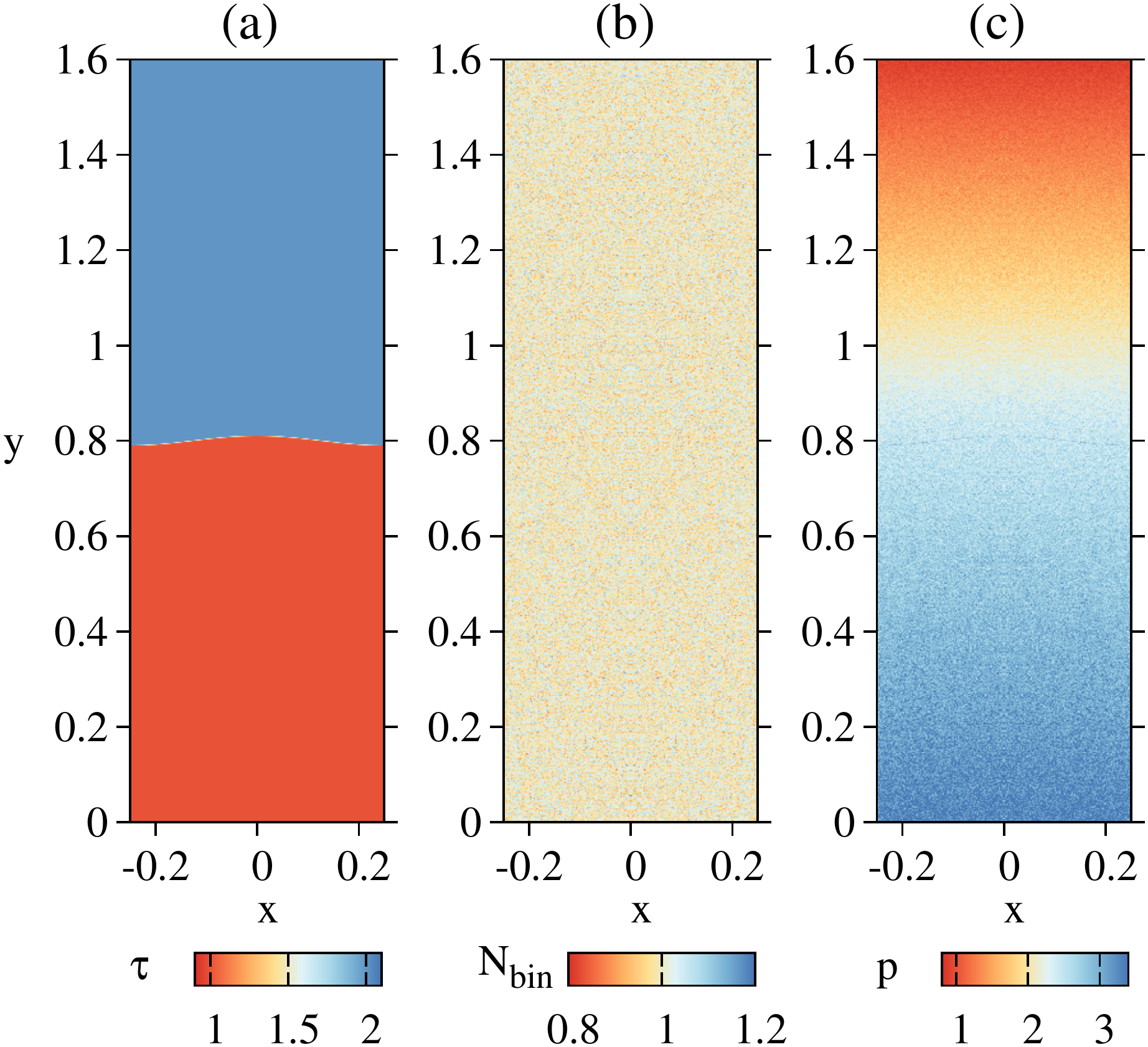}
\caption{(Color online) Initialization of smRTIs with (a) average particle type $\tau$, (b) average particle number $N_\mathrm{bin}$ in units of $N_0 = 625$, and (c) average pressure $p$.}
\label{RT_I}
\end{center}
\end{figure}
\begin{figure}
\begin{center}
\includegraphics[width = 0.42\textwidth]{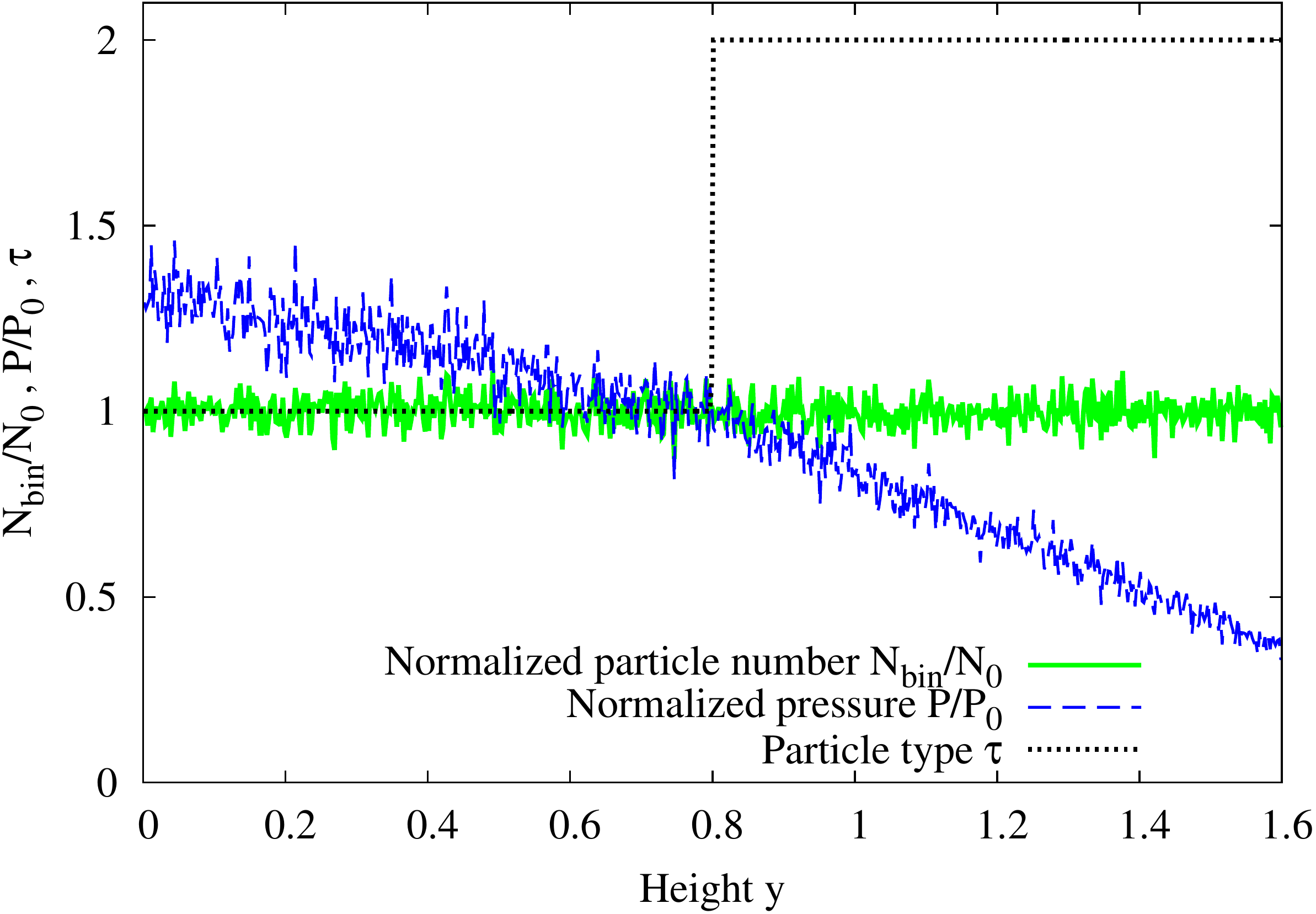}
\caption{(Color online) Profiles of the initial average particle density $N_\mathrm{bin}/N_0$, average particle pressure $P/P_0$, and particle type $\tau$ taken at $x=0.125$.}
\label{prof_init}
\end{center}
\end{figure}
Fig.\:\ref{RT_I} shows the initial (a) particle type $\tau$, (b) normalized particle number with $N_0 = N/(L_x \times L_y) = 625$, and (c) pressure $P$. All quantities are given as averages per output bin and we mirror the results at $x=0$. Figure\:\ref{prof_init} provides an estimate on the level of statistical noise in the simulation via y-profiles of the density, pressure, and particle type taken at $x=0.125$. We see significant fluctuations in $N_\mathrm{bin}$ and $P$ and, as a consequence, will average output quantities over several output bins in our later analysis. The simulation evolves up to $t = 3.75$ and the results for $\tau$ are plotted in Fig.~\ref{RT_mfp} for (a1) $t = 0.5$, (a2) $t=1.25$, (a3) $t=1.75$, (a4) $t=2.5$, and  (a5) $t=3.75$. For $t \leq 1.75$ we add the analytic solution from linear theory using $l = l_\mathrm{min,1}$ and $l = l_\mathrm{min,2}$ as dashed and solid lines, respectively. For better visualization, the output is limited to $0.2 \leq y \leq 1.4$.\\
As before, we see the formation of a diffusion layer for $t \leq 0.5$ caused by the finite number of test-particles and the finite value of $l$. At $t=0.5$, the smRTI amplitude is in good agreement with the analytic prediction but small perturbations are present and serve as seeds for fluid instabilities which become visible at $t \sim 1.25$. Here, in addition to the growth of the smRTI, small bubbles of light fluid are visible. Overall, the analytic prediction with $l_\mathrm{min,1}$ and $l_\mathrm{min,2}$ match the envelope of the bubble maxima. However, at $t \geq 1.75$, the amplitude of the smRTI significantly lags behind the prediction with $l_\mathrm{min,1}$ while $l_\mathrm{min,2}$ provides a better fit. The disagreement with $l_\mathrm{min,1}$ might indicate that the minimal mean-free-path is indeed given by $l_\mathrm{min,2}$. The smRTI evolution could also be affected by the secondary instabilities. These are clearly seen for $t \geq 1.75$ and might make comparisons with analytic predictions less reliable. Furthermore, a finite mean-free-path introduces compressibility effects which have been shown to change the RTI evolution \cite{Wei12}. To estimate their impact, we calculate the Mach number \cite{Reckinger10}:
\begin{equation}
M = \sqrt{(\rho_1 + \rho_2) g \lambda/(2P_{0})}\:. 
\end{equation}
Our simulations have $M \sim 0.55$ and are therefore in the compressible subsonic flow regime (with $M \lesssim 0.3$ for incompressible subsonic flow). Non-viscous, ideal fluids have a growth rate of $\gamma_i \sim 2.046$ resulting in $\Gamma_i = 17.960$ for $t=1.75$. When compressibility is included, $\gamma$ decreases to $\gamma_{c} \sim 2.01$ (we use eq.(31) of \cite{Livescu04}), and the corresponding amplitude gain is $\Gamma_{c} (t=1.75) = 16.865$. The difference between $\Gamma_i$ and $\Gamma_{c}$ is small and the amplitude reduction $\Delta \eta$ for $x=0$ and $t=1.75$ is
\begin{align}
\Delta \eta (0,1.75) &=  \eta_{i} (0,1.75) - \eta_{c} (0,1.75) \nonumber\\
& \sim  0.980 - 0.969 = 0.011\: .
\label{delta_eta}
\end{align}
This is only about $1.1$\% of the smRTI amplitude and thereby too small to account for the disagreement seen in Fig.\ref{RT_mfp}(a3). Of course, the impact of compressibility could be larger when viscosity and diffusion are taken into account. We will come back to this point in the next section.
\subsection{Mean-free-path comparison}
\label{subsection:mfp}
\begin{figure}
\centering
\includegraphics[width=0.42\textwidth]{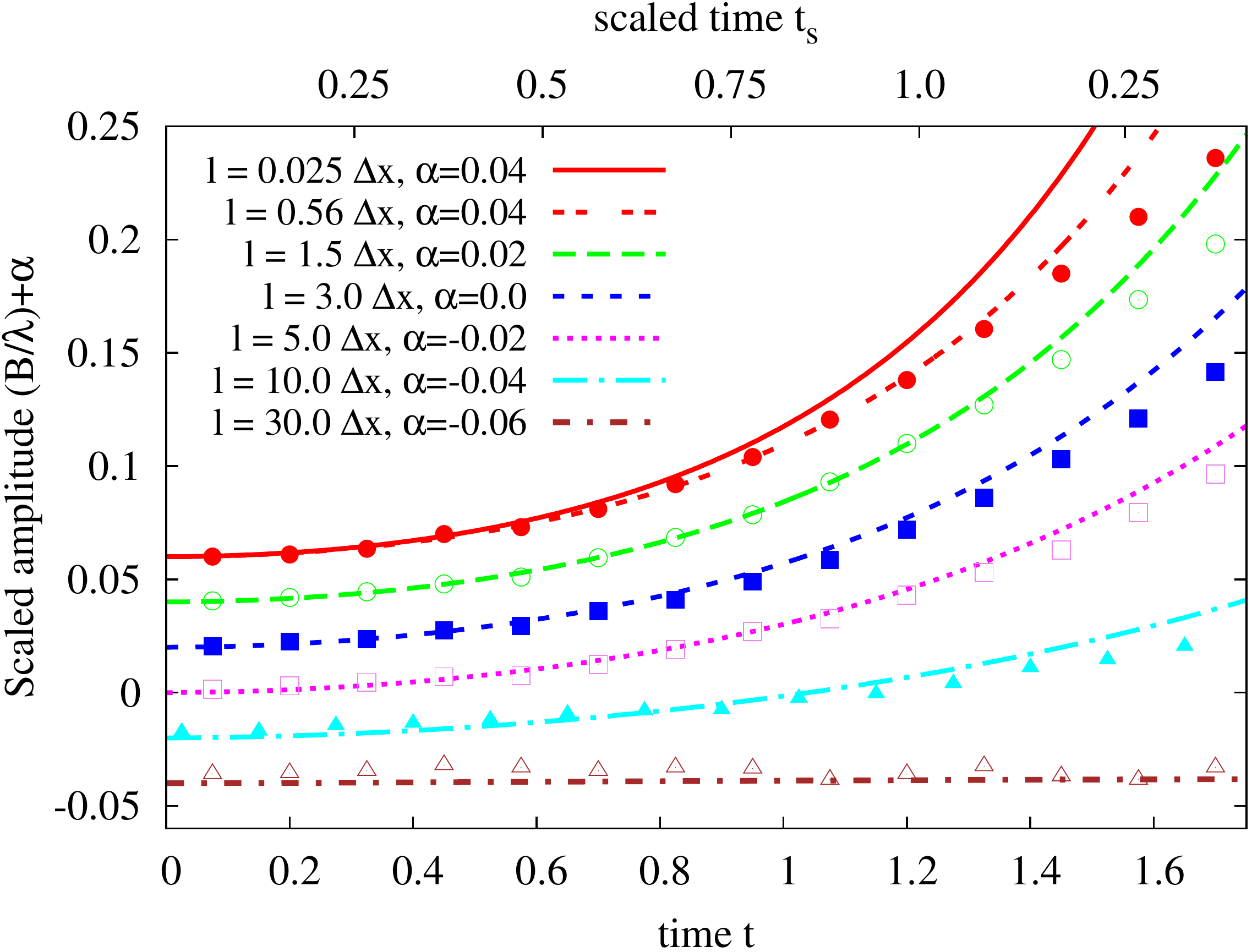}
\caption{(Color online) Time evolution of the smRTI amplitude $B$ for $l = 0.025 \: \Delta x$ (red filled circles), $l = 1.5 \: \Delta x$ (green empty circles),  $l = 3.0 \: \Delta x$ (blue filled squares), $l = 5.0 \: \Delta x$ (pink empty squares), $l = 10.0 \: \Delta x$ (cyan filled triangles), and $l = 30.0 \: \Delta x$ (brown empty triangles) scaled by the perturbation wavelength $\lambda$ together with the linear theory solution (lines).}
\label{amp}
\end{figure}

\begin{figure}
\begin{center}
\includegraphics[width = 0.42\textwidth]{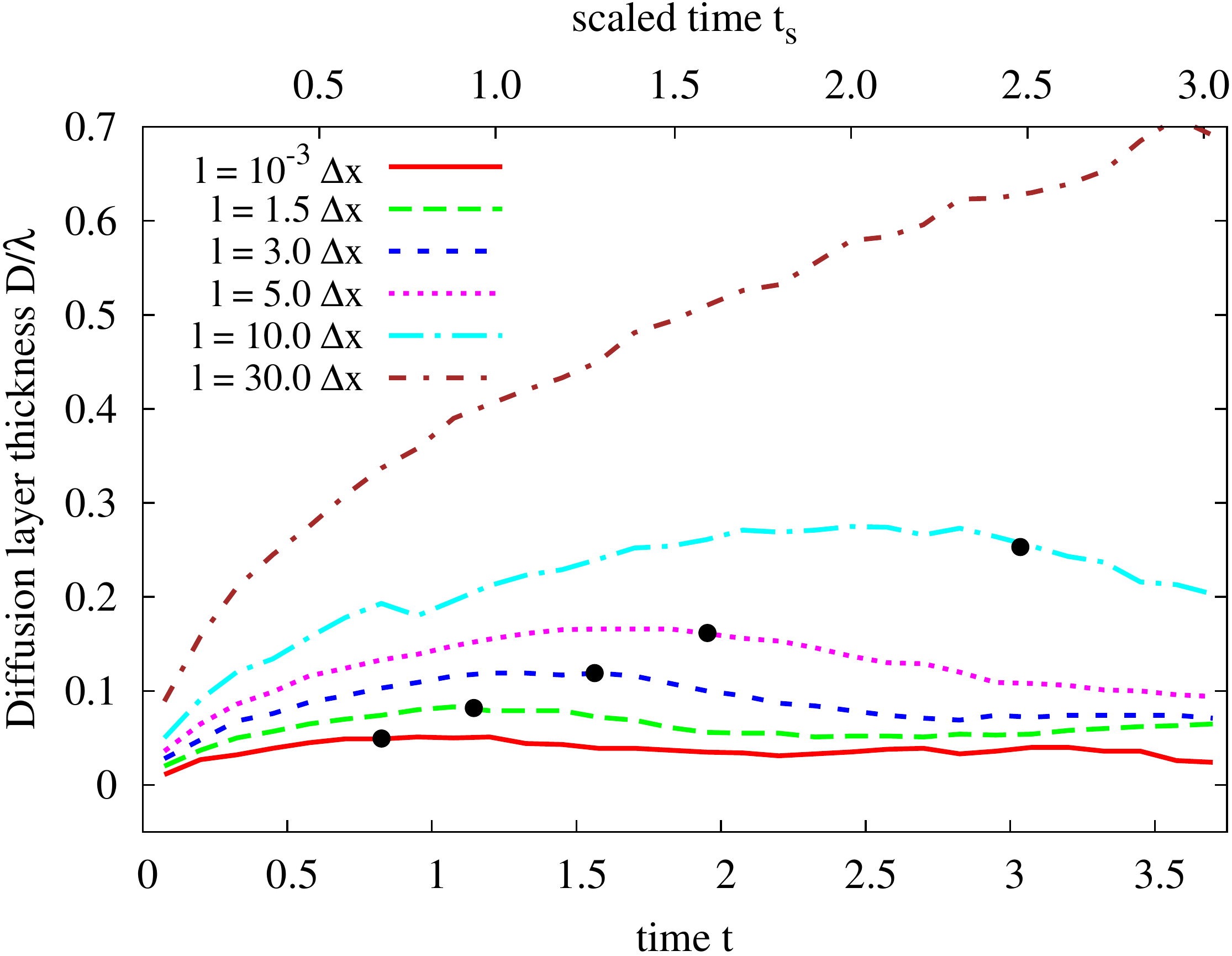}
\caption{(Color online) Diffusion layer widths of smRTI simulations scaled by the perturbation wavelength $\lambda$ as a function of time. Black dots correspond to points when $B > D$.}
\label{diff_layer}
\end{center}
\end{figure}

Figures\:\ref{RT_mfp}\:(b)-(g) show the smRTI for (b) $l = 1.5\:\Delta x$, (c) $l = 3.0\:\Delta x$, (d) $l = 5.0\:\Delta x$, (e) $l = 10.0\:\Delta x$, (f) $l = 30.0\:\Delta x$, and (g) $l = 100.0\:\Delta x$ whereas subfigures (1)-(5) correspond again to different time snapshots as described in the previous section. The jump from $l = 0.025\:\Delta x$ to $l = 1.5\:\Delta x$ is motivated by a previous finding that shock wave dynamics do not differ significantly for $l \lesssim \Delta x$ \cite{Sagert14}. We expect a similar outcome for the smRTI simulation. Of course, $l$ still impacts particle diffusion, compressibility and viscosity \cite{Wei12}, however, we expect stronger effects for $l > \Delta x$. For comparison with the simulations, we use again the viscous diffusive growth rates $\Gamma(t)$ from table \ref{table1} and plot the analytical predictions for $t \leq 1.75$ in Fig.\:\ref{RT_mfp}. It can be seen that the latter agree with the simulations. One possible explanation for the lag of the smRTI with $l=0.025\:\Delta x$ at $t=1.75$ was the effect of finite compressibility. However, the general agreement between theory and simulations for $1.5\: \Delta x \leq l \leq 5 \: \Delta x$ indicates that compressibility does not play a large role for the smRTI  height. The lag is therefore either caused by secondary instabilities or indicates that the true minimal mean-free-path is given by $l_\mathrm{min,2}$.\\
\newline
Fig.\:\ref{RT_mfp} also demonstrates the growth of the mixed fluid layer thickness with $l$ as well as the accompanying blurring of small-scale perturbations. While the simulation with $l = 30.0 \: \Delta x$ still exhibits signs of a weak smRTI development, no RT evolution can be seen for $l = 100.0 \: \Delta x$ and both fluids simply mix over time as was discussed in section \ref{section:gamma_D_nu}. For such large values of the mean-free-path, the simulation approaches to the regime of non-interacting gas. \\
For small $l$, the average position of the mixing region matches the analytic prediction for $t \lesssim 1.75$. We can also see the impact of the mean-free-path on the smRTI amplitude in Fig.\:\ref{RT_mfp}\:(3), where the latter clearly decreases with larger $l$. For a more quantitative analysis, we plot the amplitude $B$ as a function of time  $t \leq 1.75$ and scaled time $t_s = t \sqrt{Ag/\lambda}$ in Fig.\:\ref{amp}. Here, $B$ is extracted in the following way: Data is taken from the simulation every $\Delta t = 0.25$ resulting in 75 output files. For each time snapshot, we find the highest point $B_\mathrm{max}$ for $0.005 \leq x \leq 0.125$ and $\tau \leq 1.8$. Once $B_\mathrm{max}$ is determined, we pick the lowest point $B_\mathrm{min}$ with the same $x$-coordinate as $B_\mathrm{max}$ but $\tau > 1.2$. The height of the smRTI is then given by the average:
\begin{equation}
B = 0.5 \: ( B_\mathrm{max} + B_\mathrm{min} ) - 0.8 \: .
\end{equation}   
Furthermore, to suppress statistical noise, we average $B$ over five consecutive snapshots and scale it with $\lambda$. The simulation data is plotted as points while the linear theory predictions are represented by lines. To better distinguish between the different curves, we shift the data sets by a factor $\alpha$ along the $y$-axis.
Although the simulation data scatters around the analytic solutions, both generally fit very well at early times. At larger $t$, especially for $l = 0.025\: \Delta x$ the instability amplitude increases slower than the analytic prediction with $l=l_\mathrm{min,1}$ while a better fit is again achieved for $l=l_\mathrm{min,2}$.\\
\newline
Next, we compare the thickness of the diffusion layer which is determined by:
\begin{equation}
D = ( B_\mathrm{max} - B_\mathrm{min} ) .
\end{equation}   
For the output in Fig.\:\ref{diff_layer}, we scale again by $\lambda$ and see an increase of $D$ with $l$. Furthermore, we find that $D > B$ for early times which implies the domination of diffusion over smRTI growth.
The black dots in Fig.\:\ref{diff_layer} mark the times, when the latter takes over. While up to this point the width of the mixed fluid layer increases, its growth saturates and even decreases for larger $t$. This can be caused by displacement of heavy fluid at the top of the light fluid bubble due to its upward motion. Alternatively, the bubble might squeeze matter in the mixed fluid layer due to finite compressibility. Since the decrease is more pronounced for larger $l$, a cause involving finite matter compressibility seems to be more likely.
\subsection{Final state comparison}
\begin{figure*}
\begin{center}
\includegraphics[width = 0.9\textwidth]{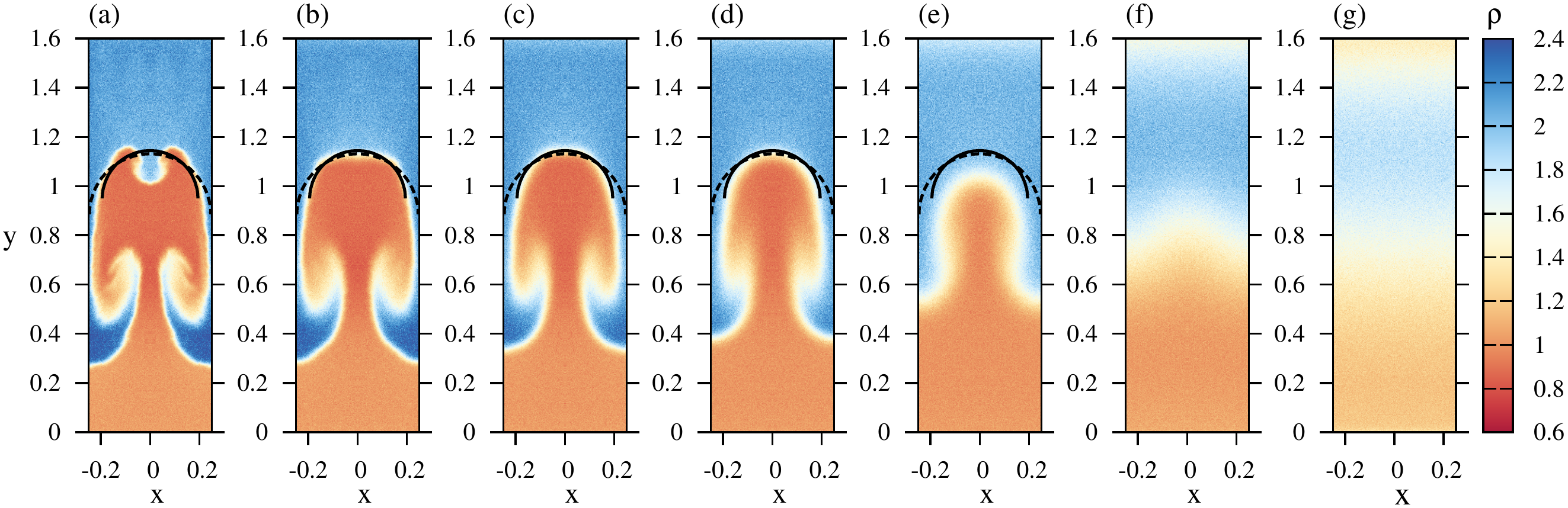}
\caption{(Color online) Normalized particle density $\rho$ via eq.(\ref{rho}) for smRTI simulations at $t=3.75$ and different mean-free-paths $l$ (subfigures (a)-(g)) as in Fig.\:\ref{RT_mfp}(5). Solid and dashed lines correspond to theoretical estimates for the bubble radius with $R_a = 0.194$ (solid) and $R = 0.258$ (dashed) in eq.(\ref{eta21}) and eq.(\ref{eta22}), respectively. See text for discussion}.
\label{RT_comp}
\end{center}
\end{figure*}

We now analyze the smRTI states at $t=3.75$ and compare them with theoretical predictions. First, we plot the normalized particle density:
\begin{equation}
\rho = m_{1,2} \: n / (m_1 N_0),\: m_1 = 10^{-8}, \: m_2 = 2 \: m_1
\label{rho}
\end{equation} 
in Fig.\:\ref{RT_comp} with subfigures (a)-(g) as in Fig.\:\ref{RT_mfp}. 
We can see two interesting phenomena: First, with larger $l$, particle densities at the top of the simulation box decrease over time. The effect is visible for $l = 30.0 \: \Delta x$ and $l = 100.0 \: \Delta x$ and is caused by the absence of scattering. Initially, particle velocities are set up according to MB distributions. Over time, the gravitational acceleration increases components in the negative $y$-direction. Due to the absence of particle interaction, the latter cannot transfer vertical velocity components into other directions, and, as a consequence, particles are accelerated downwards. Since the absolute particle velocity decrease with height, the effect is most pronounced at the top of the simulation box. The second effect is present for $l \leq 3.0 \: \Delta x$ where the density of the heavy fluid at the foot of the smRTI is increased in comparison to the top of the simulation space. This could be attributed to compressibility as descending spikes squeeze matter when they move downwards. Different from the diffusion layer width, the effect is most pronounced for small mean-free-paths. It could be caused larger spike velocities for small $l$ which result in stronger matter compression.\\
\begin{figure}
\begin{center}
\includegraphics[width = 0.42\textwidth]{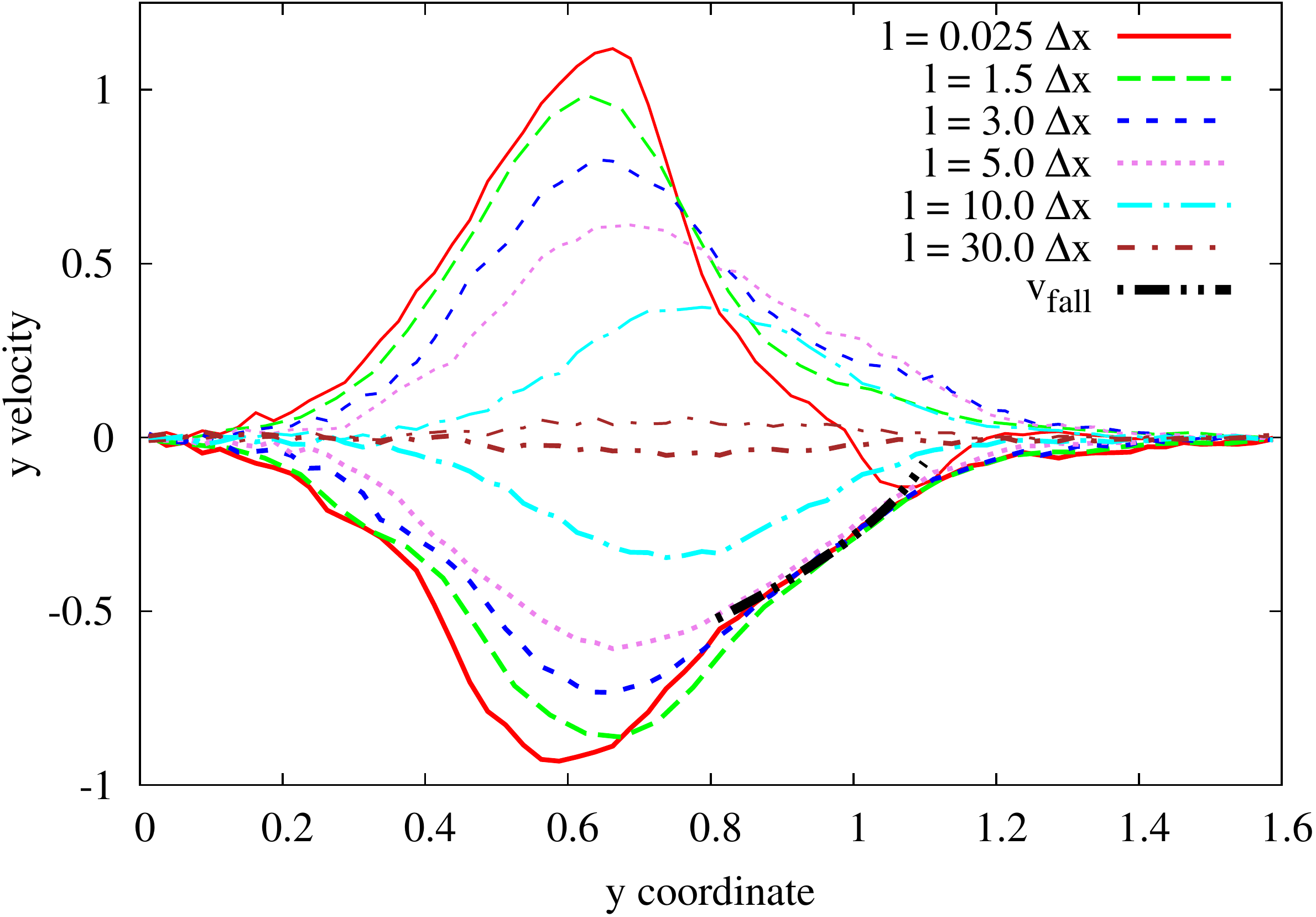}
\caption{(Color online) Velocity profiles of the rising smRTI bubble (thin lines) and sinking spike (thick lines) for different mean-free-paths $l$ along the y-axis at $t = 3.5$. The velocities of the sinking material are compared to the free fall velocity \cite{Inogamov99} (thick dash-dot-dot line).}
\label{vely}
\end{center}
\end{figure}
To test this assumption, we determine vertical velocity profiles of the rising bubble and the sinking spike. These are shown in Fig.\:\ref{vely}, where we plot the $y$-velocities per bin averaged over $0.005 \leq x \leq 0.015$ for the rising bubble and $0.235 \leq x \leq 0.245$ for the descending spike. Furthermore, we average the obtained velocities over 10 consecutive output bins in the $y$-direction. The resulting profile shapes generally agree with expectations \cite{Inogamov99} whereas we find a clear dependence on $l$. The negative bubble velocity for $l = 0.025\:\Delta x$ at $1.0 \leq y \leq 1.2$ is caused by the secondary instability at its top. The largest absolute velocities are found in the $l = 0.025 \: \Delta x$ case which supports our assumption that the stronger compression of matter occurs due to larger spike velocities for small mean-free-paths. Starting at the apex of the light fluid bubble, the horizontal component of the spike velocity should be small and the particle motion dominated by the vertical downward component \cite{Inogamov99}. As a consequence, we can compare the spike velocities to the free-fall velocity of the heavy fluid in the gravitational field \cite{Inogamov99}: 
\begin{align}
v_\mathrm{fall} (y) &= - \sqrt{2 g \left( 1 - (\rho_1/\rho_2)  \right) (y_b - y)} + v_F, 
\label{free_fall}\\
v_F &= 0.59 \sqrt{ g \left( 1 - (\rho_1/\rho_2) \right)/k} ,
\nonumber
\end{align}
whereas $y_b = 1.132$ marks the average height of the bubble apex for $l \leq 5.0 \: \Delta x$. We find that the spike velocities in simulations with $l \leq 5.0 \: \Delta x$ agree with $v_\mathrm{fall}(y)$ for $0.8 \lesssim y \lesssim 1.1$. The deviations from eq.(\ref{free_fall}) in the lower spike region are most likely caused by the influence of the mushroom. For larger mean-free-paths, the velocity profiles become less pronounced. This is due to the larger viscosity and broader spike regions which allow particles to move horizontally in addition to their vertical motion.\\
\newline
In addition, we can also determine the bubble velocity and compare it with theoretical predictions for its asymptotic value at $t \rightarrow \infty$ \cite{Goncharov02}:
\begin{align}
& v_\mathrm{bubble,a} = 1.025 \sqrt{ \frac{2A}{A+1} \frac{g}{3k}}, \\
&\rightarrow  v_\mathrm{bubble,a}/\sqrt{g/k} \sim 0.41,
\end{align}
whereas \cite{Inogamov99} gives $v_\mathrm{bubble,b}/\sqrt{g/k} \sim 0.59$ (interestingly, the two expressions differ by a factor of $\sqrt{2}$). 
\begin{figure}
\begin{center}
\includegraphics[width = 0.42\textwidth]{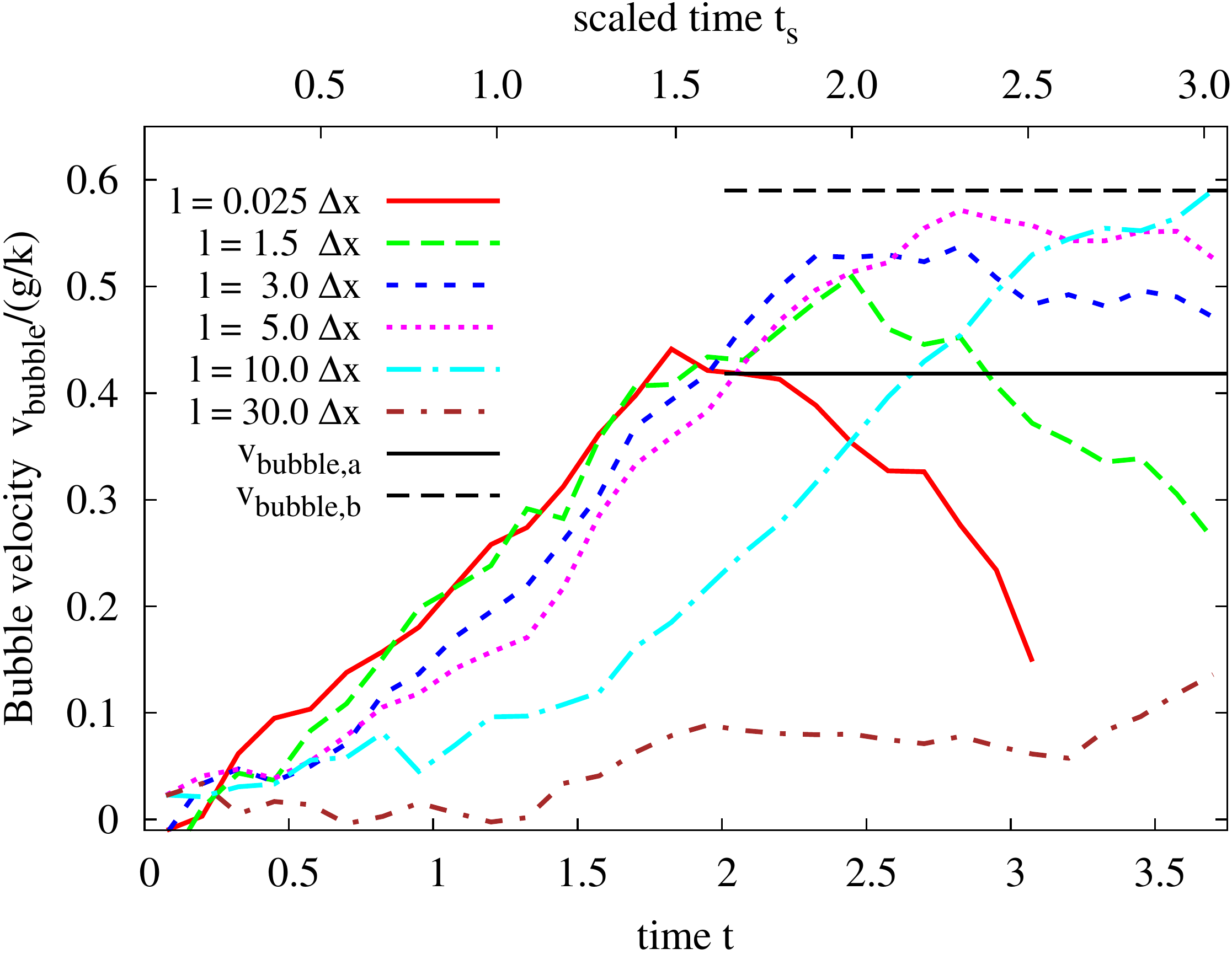}
\caption{(Color online) Scaled velocity of the smRTI bubble as a function of time for different mean-free-paths $l$. The horizontal solid and dashed lines are theoretical estimates on the asymptotic bubble velocity \cite{Goncharov02, Inogamov99} (see text).}
\label{v_bubble}
\end{center}
\end{figure}
To extract $v_\mathrm{bubble}$ in our simulations, we average all $y$-velocities per bin between $B_\mathrm{min} \leq y \leq B_\mathrm{max}$. The latter are determined for $0.005 \leq x \leq 0.015$. To further reduce statistical noise, we average $v_\mathrm{bubble}$ over five consecutive time snapshots. Fig.\:\ref{v_bubble} shows the simulation results for $v_o$ together with the asymptotic predictions for $v_\mathrm{bubble,a}$ and $v_\mathrm{bubble,b}$. For $l = 10 \: \Delta x$ and $l = 30 \: \Delta x$, the bubble velocity rises very slowly. While it eventually reaches $v_\mathrm{bubble,a}$ for $l = 10 \: \Delta x$, it stays small for $l = 30 \: \Delta x$ due to the large particle diffusion. For $l \leq 5 \: \Delta x$, we initially see a linear increase of $v_\mathrm{bubble}$ with time. The velocities eventually reach a maximum between $v_\mathrm{bubble,a} < v_\mathrm{bubble} < v_\mathrm{bubble,b}$ with a subsequent drop. The latter is more pronounced for smaller mean-free-paths and, for $l = 0.025 \: \Delta x$, is most likely caused by the secondary instability forming at the top of the bubble (see Fig.\:\ref{RT_mfp}\:(a4)). For $l = 1.5\:\Delta x$ and  $3.0\:\Delta x$ we find a similar stagnation of $v_\mathrm{bubble}$. Although here, a secondary instability is not directly visible, the flat bubble top in Fig.\:\ref{RT_mfp}\:(b5) and (c5) could be interpreted as its onset which decreases $v_\mathrm{bubble}$. The effect is very weak for $l=5.0 \: \Delta x$. Here, the top of the bubble is round and the velocity does not exhibit a large decrease with time.\\
\newline
Another effect that needs to be addressed in the future is the size of the simulation space. As the light fluid bubble rises, it approaches the upper boundary of the simulation space. Although we apply random reflective boundary conditions, wave reflection might still occur and, when propagating downwards, could interact with the bubble and cause it to deform or decelerate. In addition, as the bubble rises and expands, it comes very close to the vertical box boundaries. These can impact the evolution of the smRTI by preventing its full expansion. To explore the latter, we compare the radius of the light fluid bubble with theoretical predictions. The 2D asymptotic value of the bubble curvature $\kappa$ was determined by \cite{Goncharov02} for an ideal fluid as:
\begin{equation}
\kappa_a  = - 2 \pi/ (4.88 \lambda) \sim - 2.575,
\label{eta21}
\end{equation}
with a corresponding radius $R_a = -1/(2 \: \kappa_2) \sim 0.194$ and by \cite{Abarzhi03} (using Fig.\:1 of the reference with $\zeta_1/k \sim -0.105$) to:
\begin{equation}
\kappa_b \sim - 0.105 (16 \pi/ (3 \sqrt{3} \lambda)) \sim - 2.031.
\label{eta22}
\end{equation}
The latter results in $R_b \sim 0.246$. It is noteworthy that $R_b \sim L_x$ which might lead to wall-effects in the late stages of the smRTI. We plot semi-circles with $R_a$ (solid line) and $R_b$ (dashed line) together with the particle densities of the smRTIs in Fig.\:\ref{RT_comp}. Since $R_a$ and $R_b$ are derived for ideal fluids, we expect the best fit for close-to-continuum simulations, while both radii should overestimate the bubble size for calculations with large mean-free-paths. This effect is clearly seen for $l = 10.0 \: \Delta x$. However, $R_a$ seems to reproduce the smRTI with $l \sim 5.0 \: \Delta x$ best, while underestimating the bubble for smaller mean-free-paths. This could imply that $R_b$ is a better estimate on the radius, however, when comparing with numerical results we find that $R_b$ seems to always overestimate the smRTI bubble, even for $l = 0.025 \: \Delta x$. This could be due to the small width of the simulation space which might restrict the bubble and result in a smaller radius. Studies of the smRTI in a bigger simulation space need to address the bubble evolution in the future.
\subsection{smRTI with small mean-free-paths}
\label{section:small_l}
\begin{figure}
\begin{center}
\includegraphics[width = 0.45\textwidth]{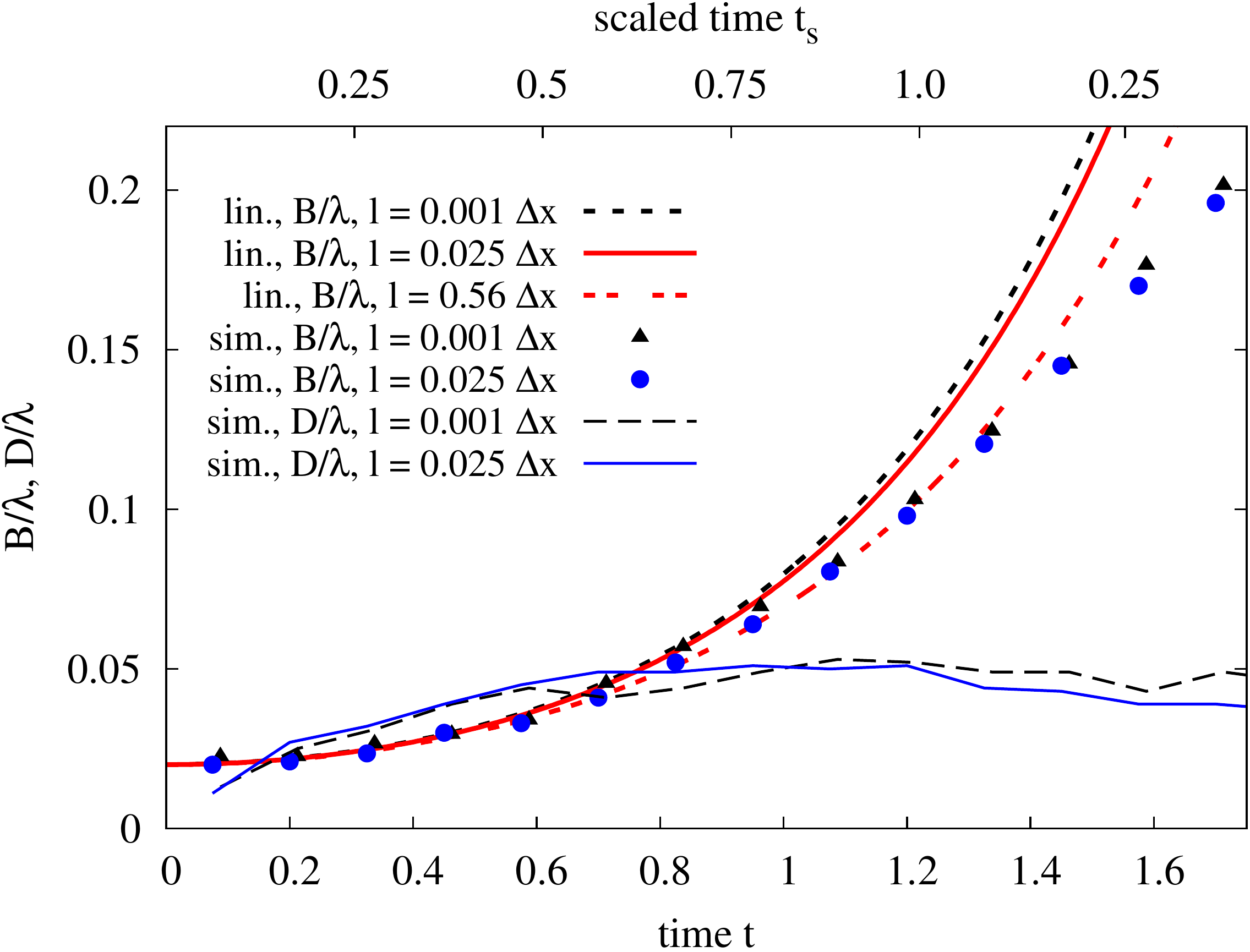}
\caption{(Color online) Scaled amplitude $B/\lambda$ (filled triangles and circles) for $l = 10^{-3} \: \Delta x$ and $l = 0.025 \: \Delta x$ together with the linear theory predictions (thick lines) for $l = 10^{-3} \: \Delta x$,  $l = 0.025 \: \Delta x$ and  $l = 0.56 \: \Delta x$. Simulation results for the diffusion layer width $D/\lambda$ for $l = 10^{-3} \: \Delta x$ and $l = 0.025 \: \Delta x$ is given by thin dashed and solid lines, respectively.}
\label{min_comp}
\end{center}
\end{figure}

As mentioned before, the particle mean-free-path in our studies is limited by a minimal value $l_\mathrm{min}$. Considering that collision partners cannot be farther away than $\sim 2 \: \Delta x$, this value should be given by $l_\mathrm{min,1} = 0.025 \: \Delta x$. On the other hand, when assuming that for large $N_\mathrm{bin}$, collision partners are typically close to each other and using an average distance of $2 \sqrt{ A_\mathrm{bin}/(\pi N_\mathrm{bin})}$, the minimal mean-free-path increases to $l_\mathrm{min,2} \sim 0.56 \: \Delta x$ (see discussion in section \ref{section:minimal_l}). Previously, we argued that simulations should evolve similarly for $l \leq l_\mathrm{min}$, especially properties like diffusion layer width and smRTI amplitude - quantities that directly depend on $l$ - should not differ much. We will test this assumption by setting $l \ll \Delta x =10^{-3} \: \Delta x$ and performing a smRTI simulation. \\ 
\begin{figure*}
\begin{center}
\includegraphics[width = 0.75\textwidth]{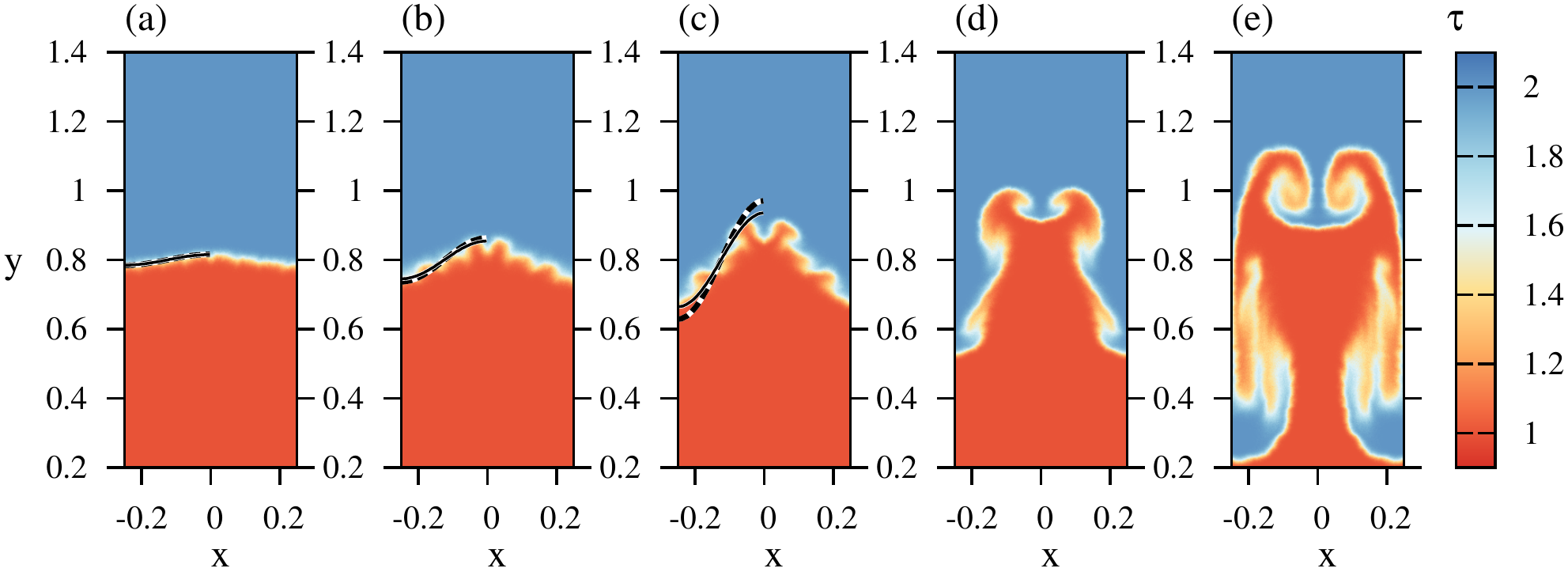}
\caption{(Color online) Particle type $\tau$ for the smRTI with diffusion suppression, mean-free-path $l = 10^{-3} \: \Delta x$ and particle number $N = 4.0 \times 10^7$. Subfigures (a)-(e) correspond to different simulation times as in Fig.\:\ref{RT_mfp}. Dashed lines give the analytic prediction for the smRTI amplitude for $l = l_\mathrm{min,1} = 0.025 \: \Delta x$ while solid lines were obtained using $l = l_\mathrm{min,2} = 0.56 \: \Delta x$.}
\label{RT_small}
\end{center}
\end{figure*}
Fig.\:\ref{RT_small} shows the corresponding simulation snapshots whereas subfigures (a)-(g) are as in Fig.\:\ref{RT_mfp}. While the diffusion layer width is very similar to $l= 0.025 \: \Delta x$, large-wavenumber instabilities seem to be more pronounced. This is clearly visible in subfigures (b) and (c). Furthermore, in the latter, the upper small-scale instability develops into a RTI itself. The corresponding small fluid bubble moves upwards together with the smRTI. As it approaches the simulation walls, the lower part of the bubble is deflected downwards while the upper part continues to move upwards, distorting the mushroom shape. \\
For a quantitative comparison of the smRTI amplitude and diffusion layer width, we determine both as in section \ref{subsection:mfp} and plot them together with the simulation result for $l = 0.025 \: \Delta x$ and linear theory predictions for $l= 10^{-3} \: \Delta x$, $l= 0.025 \: \Delta x$, and $l= 0.56 \: \Delta x$ in Fig.\:\ref{min_comp}. The formation of secondary instabilities for $l= 10^{-3} \: \Delta x$ complicates a reliable determination of the smRTI amplitude and, as a consequence, we limit the comparison to $t \leq 1.75$.  
As we can see, the linear theory solution predicts a slightly higher amplitude for $l= 10^{-3} \: \Delta x$ in comparison to $l= 0.025 \: \Delta x$. The values for $B/\lambda$ in the simulations, on the other hand, are very similar. The same applies to the diffusion layer widths. This confirms our prediction that mean-free-path dependent quantities, such as smRTI amplitude and mixed layer width, are given by the true minimal value of $l_\mathrm{min}$. The more pronounced small-scale structures for $l = 10^{-3} \: \Delta x$ might be caused by a different sequence of random numbers in the simulation. For close-to-continuum simulations, small differences at early times could lead to different perturbations of the fluid interface and result in different smRTI evolutions. We are working on an algorithm that will suppress particle diffusion at early times and thereby suppress the evolution of small-scale structures.
\section{Summary and Outlook}
\label{section:conclusion}
We present simulations of single-mode Rayleigh Taylor instabilities (smRTIs) with a large-scale modified Direct Simulation Monte Carlo Code (mDSMC). Our approach combines the computational scaling of DSMC methods and the spatial accuracy of the Point-of-Closest-Approach technique. The aim of the current work is to test our kinetic code on its ability to reproduce fluid instabilities and study the latter for finite viscosity and diffusion. The code has been validated in the hydrodynamic regime by 2D and 3D shock wave studies in the past and is able to simulate matter for a large range of Knudsen numbers. For our RTI simulations, we apply $N= 4.0 \times 10^7$ test-particles. At early stages of the smRTI, the growth rate can be analytically obtained from linear theory while for late times the onset of secondary instabilities and turbulent mixing is seen by hydrodynamic codes and experiments. We compare our simulations to the expected behavior of the smRTI in these regimes. Our kinetic method is limited by a minimal value for the particle mean-free-path $l$, which depends on the particle number per simulation cell. Applying different $l$ we can simulate matter in a regime that is close to the continuum limit and for non-equilibrium matter. For $l \lesssim 5 \Delta x$, we find the characteristic mushroom shape of the smRTI, which is observed in hydrodynamic simulations and experiments. Furthermore, in our close-to-continuum simulations, initial irregularities of the fluid interface result in the formation of large wavenumber instabilities which evolve into RTIs themselves. A diffusion layer, caused by particles moving form one fluid into the other, is always present. Its width increases for larger $l$ while small-scale structures become blurred out. For large mean-free-paths, simulations eventually approach the regime of non-interacting gas.\\
Overall, our simulations agree with the analytic prediction from linear theory including diffusion and viscosity and lead to similar smRTI shapes as we would expect from hydrodynamic studies. We conclude that our kinetic approach can reproduce the general features of smRTI. In the future, we plan to perform convergence tests with a larger number of test-particles and thereby smaller mean-free-paths as well as more general fluid instability studies. With that, together with the already successfully passed shock wave tests, we will be able to point our attention to the simulation of e.g. astrophysical systems, like core-collapse supernovae.
\section{Acknowledgements}
This work used the Extreme Science and Engineering Discovery Environment (XSEDE), 
which is supported by National Science Foundation grant number OCI-1053575. Furthermore, 
I.S. acknowledges the support of the High Performance Computer Center and the Institute for 
Cyber-Enabled Research at Michigan State University. This research was supported in part by 
Lilly Endowment, Inc., through its support for the Indiana University Pervasive Technology Institute, 
and in part by the Indiana METACyt Initiative. The Indiana METACyt Initiative at IU is also 
supported in part by Lilly Endowment, Inc. Furthermore, the authors would like to thank LANL 
physicists Daniel Livescu and Wesley Even for useful discussion. 
\bibliography{references}

\end{document}